\documentclass[twocolumn,prl,english]{revtex4-1}
\usepackage[T1]{fontenc}
\usepackage[latin9]{inputenc}
\usepackage{verbatim}
\usepackage{amsmath}
\usepackage{amssymb}
\usepackage{graphicx}
\usepackage{babel}
\usepackage{bm}
\usepackage[normalem]{ulem}

\newcommand{\ubar}[1]{\mkern 1.5mu{\uline{{\mkern-1.5mu#1\mkern-1.5mu}}}\mkern 1.5mu}
\newcommand{\ubaro}[1]{\mkern 1.5mu\mbox{\uline{{$\mkern-1.5mu#1\mkern-1.5mu$}}}\mkern 1.5mu}
\newcommand{\uubar}[1]{\ubar{\ubaro{#1}}}
\newcommand{\uuubar}[1]{\ubar{{\ubar{\ubaro{#1}}}}}

\newcommand\uuuline{\bgroup\markoverwith%
	{%
		\textcolor{black}{\rule[-0.5ex]{2pt}{0.4pt}}%
		\llap{\textcolor{black}{\rule[-0.8ex]{2pt}{0.4pt}}}%
		\llap{\textcolor{black}{\rule[-1.15ex]{2pt}{0.4pt}}}%
	}%
	\ULon}

\makeatother

\makeatletter

\usepackage{xcolor}
\newcommand*\diff{\mathop{}\!\mathrm{d}}

\begin{document}
\renewcommand{\ULdepth}{1pt}
\title{Non-affine deformation of semiflexible polymer and fiber networks}

\author{Sihan Chen$^{1,2}$, Tomer Markovich$^{2,3,4}$, Fred C. MacKintosh$^{1,2,5,6}$}
\affiliation{$^1$Department of Physics and Astronomy, Rice University, Houston, TX 77005\\
$^2$Center for Theoretical Biological Physics, Rice University, Houston, TX 77005\\
$^3$School of Mechanical Engineering, Tel Aviv University, Tel Aviv 69978, Israel\\
$^4$Center for Physics and Chemistry of Living Systems, Tel Aviv University, Tel Aviv 69978, Israel\\
$^5$Department of Chemical and Biomolecular Engineering, Rice University, Houston, TX 77005\\
$^6$Department of Chemistry, Rice University, Houston, TX 77005}

\begin{abstract}
Networks of semiflexible or stiff polymers such as most biopolymers are
known to deform inhomogeneously when sheared. The effects of such
non-affine deformation have been shown to be much stronger than for
flexible polymers. To date, our understanding of non-affinity in such systems is limited to simulations or specific 2D models of athermal fibers. Here, we present an effective medium theory (EMT) for non-affine deformation
of semiflexible polymer networks, which is general to both 2D and 3D and
in both thermal and athermal limits. The predictions of this model are
in good agreement with both prior computational and experimental results
for linear elasticity. Moreover, the framework we introduce can be
extended to address nonlinear elasticity and network dynamics. 
\end{abstract}

\maketitle
Networks of stiff or semiflexible polymers are vital for the function of most living systems. Such networks control much of the elastic properties of biomaterials ranging from the cell cytoskeleton to extracellular matrices at the tissue scale~\cite{Fletcher2010485,Shivers21037,van2019emergence,hudnut2018role}. Over the past few decades there has been significant progress in our understanding of the fundamental physical properties of semiflexible networks~\cite{MacKintosh1995,Isambert1996,kroy1996force,Gittes1998,Hinner1998,Morse19987030,Gardel2004,Storm2005191,Mizuno2007,Chaudhuri2007, stenull2011simple,Broedersz2014,pritchard2014mechanics}. Previous studies of 2D and 3D semiflexible networks have, among other things, revealed a transition from bend-dominated, non-affine regime to a stretch-dominated, affine regime~\cite{Head2003,Wihelm2003,onck2005alternative,Das2007,Broedersz2012, Shahsavari2012,Mao2013,Shahsavari2013, Nguyen2016, Licup2016, Baumgarten2021} that is governed by the average polymer length (or molecular weight), in stark contrast with flexible polymer systems. 

The classical theory of {\it rubber elasticity}~\cite{Rubinstein,James1947, James1948} is very successful in describing the elastic properties of flexible polymers networks. Early approaches assumed deformations to be affine, with uniform strain on all scales. The phantom-network model relaxed this assumption and showed that local network structure indeed affects elastic properties, but in a way that does not change the basic scaling with macroscopic quantities such as average polymer length, system volume, temperature, etc.~\cite{Flory1985, miehe2004micro,raina2014homogenization}. By contrast, the strong bending rigidity of semiflexible polymers invalidates the phantom model and leads to much stronger non-affine effects~\cite{Head2003, Wihelm2003,Head2003PRE}, including a surprising dependence on dimensionality~\cite{Broedersz2012}. Most of the prior work accounting for non-affinity in semiflexible networks has been limited to numerical simulation~\cite{onck2005alternative, palmer2008constitutive,huisman2008monte,cioroianu2016disorder,islam2018effect}, while a theory analogous to the phantom network has been lacking, especially in 3D. Various models based on effective medium theories (EMT) introduced for rigidity percolation~\cite{phillips1979topology,thorpe1983continuous,Feng1985} have been proposed for lattice-based or topologically similar networks~\cite{Das2007,broedersz2011criticality,Sheinman2012,Mao2013,Mao20132,mao2015mechanical,huang2022shear,damavandi2022effective}, along with floppy mode models for off-lattice networks~\cite{Heussinger2006,Zhou2018}. But, both of these approaches have been limited to 2D networks and have neglected important thermal fluctuations.

Here, we develop an analytical model for the elasticity of both thermal semiflexible polymer and athermal fiber networks that accounts for the non-affine deformations. Our model applies to both lattice-based and random off-lattice networks that are isotropic and homogeneous on large scales. As we show, this model can be applied to both thermal and athermal networks. Our prediction of the bend-to-stretch transition quantitatively agrees with previous athermal simulations of 3D networks, while explaining the different scaling dependences on filament length in 2D lattice and off-lattice (e.g., Mikado) networks. Moreover, for thermal networks where simulations are lacking, our model predicts a bend-to-stretch transition that agrees with previous experiments~\cite{Jaspers2014,Nguyen2016}. Although we focus here on the linear elastic limit, this model can also be extended to address the role of non-affine fluctuations in the dynamics~\cite{Tighe2012,yucht2013dynamical,Milkus2017,Shivers2022}, stress-stiffening~\cite{Gardel2004,Storm2005191} and recently identified strain-controlled criticality~\cite{Sharma2016, Vermeulen2017, merkel2019minimal, arzash2020finite}. 

We begin by considering an athermal crosslinked semiflexible polymer network in 3D. The discussion on 2D and thermal networks is postponed to later. The network is formed by $N$ filaments each with polymer length $L$ and point-like hinged crosslinkers with average crosslinking distance $\ell_c$. Its Hamiltonian is:
\begin{align}
H_O=\sum_{\alpha=1}^N \Big[H_{b}\left[{\bm u}^{\alpha}(s)\right]+ H_{s}\left[\bm u^{\alpha}(s)\right]\Big]\,,\label{e1}
\end{align}
where ${\bm u}^{\alpha}(s)={\bm u}_\parallel^{\alpha}(s)+{\bm u}_\perp^{\alpha}(s)$ is the microscopic displacement of the $\alpha$-th polymer at position $s$ along its contour ($-L/2<s<L/2$), with ${\bm u}_\parallel^{\alpha}(s)$ and  ${\bm u}_\perp^{\alpha}(s)$ being its longitudinal and transverse components, respectively.  $H_{b}\left[{\bm u}(s)\right]=\kappa \int \diff s |\partial ^2{\bm u}_\perp/\partial s^2|^2/2$ and $H_{s}\left[{\bm u}(s)\right]=\mu \int \diff s |\partial {\bm u}_\parallel/\partial s|^2/2$ are the bending and stretching energy, respectively.  If a crosslink exists between the $\alpha$-th and the $\beta$-th polymer, it leads to an additional constraint, ${\bm u}^{\alpha}(s_{\alpha\beta})={\bm u}^{\beta}(s_{\beta\alpha})$, with $s_{\alpha\beta}$ ($s_{\beta\alpha}$) being the position of the crosslink on the $\alpha$-th ($\beta$-th) polymer. 

We are interested in how this network deforms under an external shear stress $\sigma_O$. For athermal networks the deformation is found from the minimum-energy state, in which the microscopic deformations of each polymer are denoted by ${\widetilde{\bm u}}^{\alpha}(s)$ and the shear strain of the entire network is $\gamma_O$. The linear shear modulus is defined as $G_O=\partial  \sigma_O/\partial \gamma_O|_{\gamma_O=0}$. For simplicity we assume a stress $\sigma_O$ in the $x-z$ plane. This causes (in the linear regime) a simple shear of the $x-z$ plane in the $x$ direction, such that $\gamma_O$ corresponds to a single nonzero term $\Lambda_{xz}=\gamma_O$ in the deformation tensor $\uubar{\bm \Lambda}$. 

Although the network Hamiltonian has a quadratic form (Eq.~(\ref{e1})), a direct analytical solution of the minimum-energy state is challenging for two reasons: the first is the existence of the crosslinking constraints, which introduces correlations between different polymers. Therefore, their deformations ${\bm u}^{\alpha}(s)$ can not be considered as independent variables; the other is the unclear relation between the  microscopic deformations (${\bm u}^{\alpha}$) and the macroscopic deformation ($\gamma_O$) for non-affine deformations. Below we detail how we overcome these challenges. 
\begin{figure}[t]
	\centering
	\includegraphics[scale=0.30]{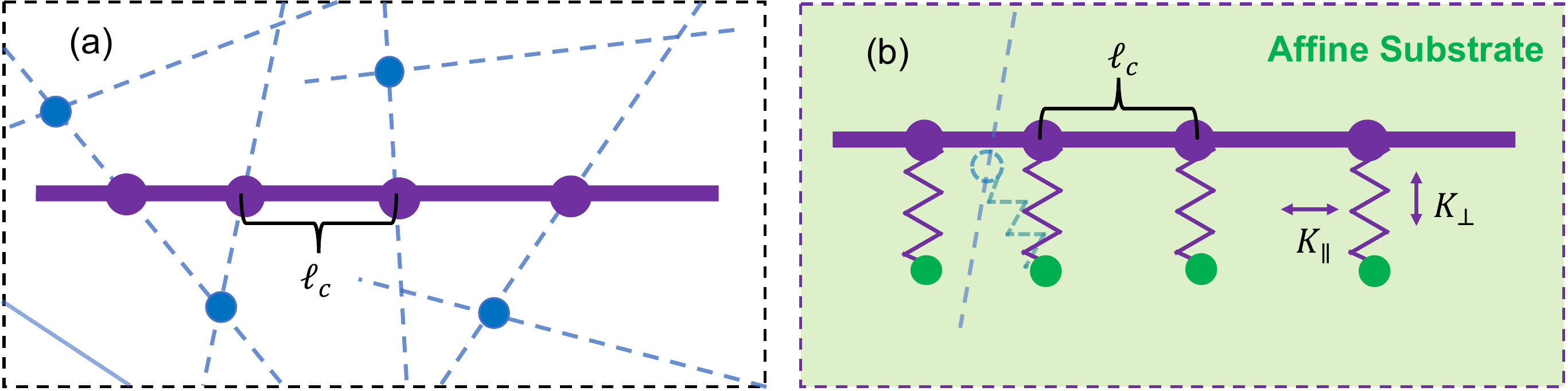}
	\caption{(a) 2D sketch of a crosslinked semiflexible polymer network in either 2D or 3D, with average crosslinking distance $\ell_c$. Each polymer has a contour length $L$. (b) Sketch of the EMT, in which crosslinks are replaced by springs that connects the polymers with a substrate which deforms affinely (two polymers connected by one crosslinker are connected to the substrate via two different springs, see, e.g., the dashed polymer).   Spring constants for the parallel and transverse directions of the connected polymer are  $K_{\parallel}$ and $K_{\perp}$, respectively. 
	}
	\label{Fig.1}
\end{figure}

To remove the crosslink constraints, we have developed an EMT, in which all the polymers in the original network are preserved while all crosslinks are removed (Fig.~\ref{Fig.1}(b)).  To mimic the restraining effect of the crosslinks, each crosslink is replaced by a spring that connects the polymer at the position of the crosslink with a substrate. The substrate can only deform affinely, and its deformation does not cost any energy. Each spring has two spring constants, $K_{\parallel}$ and $K_{\perp}$, for the parallel and transverse direction of its connected polymer, respectively. The resulting EMT has an additional elastic energy $H_K$ and the effective Hamiltonian is
\begin{align}
H_{_{\rm EM}}&=\sum_{\alpha=1}^N\Big(H_{b}\left[{\bm v}^{\alpha}(s)\right]+H_{s}\left[\bm v^{\alpha}(s)\right]+H_{K}\left[\bm v_{\rm NA}^{\alpha}(s)\right]\Big)\,,\label{e2}
\end{align}
where the microscopic deformation in the EMT is denoted by ${\bm v}^{\alpha}(s)={\bm v}_A^{\alpha}(s)+{\bm v}_{\rm NA}^{\alpha}(s)$, with ${\bm v}_A^{\alpha}(s)$ being the affine displacement and ${\bm v}_{\rm NA}^{\alpha}(s)$ being the non-affine displacement.  Note that only non-affine displacements affect $H_K$, since forces are not induced between affinely deforming polymers that simply stretch/compress uniformly. The microscopic affine displacements are given by ${\bm v}_A^{\alpha}(s)= s\uubar{\bm \Lambda} \cdot\hat{\bm n}^{\alpha}$, with $\hat{\bm n}^{\alpha}$ defining the polymer orientation. The additional energy $H_K$ is the summation of the elastic energy of all springs connected to each polymer:
\begin{align}
H_{K}\left[\bm v^{\alpha}_{\rm NA}(s)\right]&=\frac{K_\parallel}{2}\sum_{i}|\bm v^{\alpha}_{\rm NA\parallel}(s_i)|^2 \notag\\&+\frac{K_\perp}{2}\sum_{i}|\bm v^{\alpha}_{\rm NA\perp}(s_i)|^2,\label{e3}
\end{align}
where $s_i$ is the position of the $i$-th spring, and $v^{\alpha}_{\rm NA\perp}$ and $v^{\alpha}_{\rm NA\parallel}$ are the transverse and longitudinal components of $v^{\alpha}_{\rm NA}$, respectively. Importantly, terms with the  same index $\alpha$ in Eq.~(\ref{e2}) describe a single-polymer Hamiltonian in which the network structure is accounted for through a harmonic energy. Such approach is conceptually similar to the effective spring constant introduced in Refs.~\cite{Morse2001, hinsch2007quantitative}  for entangled polymer solutions, as well as  tube models for flexible polymer networks~\cite{Rubinstein2002}.

Under an imposed shear stress $\sigma_{_{\rm EM}}$, we define the microscopic deformations in the minimum-energy state of the EMT as $\widetilde {\bm v}^{\alpha}(s)$, with a shear strain  $\gamma_{_{\rm EM}}$ and an elastic modulus $G_{_{\rm EM}}= \partial \sigma_{_{\rm EM}}/\partial \gamma_{_{\rm EM}}|_{\gamma_{_{\rm EM}=0}}$. Our goal is to find an EMT that reproduces the elasticity of the original network on average, i.e. $G_{_{\rm EM}} = G_O$, the inverse of which can be rewritten using the chain rule:
\begin{align}
\sum_{\alpha i}\frac{\partial \widetilde {\bm u}^{\alpha}_{i}}{\partial \sigma_O}\cdot\frac{\partial \gamma_O}{\partial \widetilde {\bm u}^{\alpha}_{i}}=\sum_{\alpha i}\frac{\partial \widetilde {\bm v}^{\alpha}_{i}}{\partial \sigma_{_{\rm EM}}}\cdot\frac{\partial \gamma_{_{\rm EM}}}{\partial \widetilde {\bm v}^{\alpha}_{i}}\,,\label{e4}
\end{align}
where $\widetilde {\bm u}^{\alpha}_{i}=\widetilde {\bm u}^{\alpha}(s_i)$ and $\widetilde {\bm v}_{i}^\alpha=\widetilde {\bm v}^{\alpha}(s_i)$ are the displacements on the crosslink positions (symbols without tilde are arbitrary polymer displacements, while symbols with tilde denote polymer displacements in the minimum-energy state). To ensure that Eq.~(\ref{e4}) is satisfied, we look for an EMT that satisfies simultaneously
\begin{subequations}\label{e5}
\begin{equation}\label{e5a}
\left\langle\frac{\partial \widetilde {\bm u}^{\alpha}_{i}}{\partial \sigma_O}\right\rangle=\frac{\partial \widetilde {\bm v}^{\alpha}_{i}}{\partial \sigma_{_{\rm EM}}}\,,
\end{equation}
\begin{equation}\label{e5b}
\frac{\partial \gamma_O}{\partial \widetilde {\bm u}^{\alpha}_{i}}=\frac{\partial \gamma_{_{\rm EM}}}{\partial \widetilde {\bm v}^{\alpha}_{i}}\,. 
\end{equation}
\end{subequations}
In Eq.~(\ref{e5a}) we average the effects of random crosslinking angles in the original network. These requirements may not be the only appropriate ones and may appear to be stronger than necessary. However, as we will show later, this choice does lead to good agreement with the expected macroscopic elasticity. Equation~(\ref{e5a}) is essentially a coherent potential approximation (CPA) as in the classic EMT of 2D lattice-based networks~\cite{Feng1985, Supp}. Importantly, Eq.~(\ref{e5b}) is different from what is usually done in an EMT, in that it allows our EMT to deform non-affinely. 
\begin{figure}[t]
	\centering
	\includegraphics[scale=0.3]{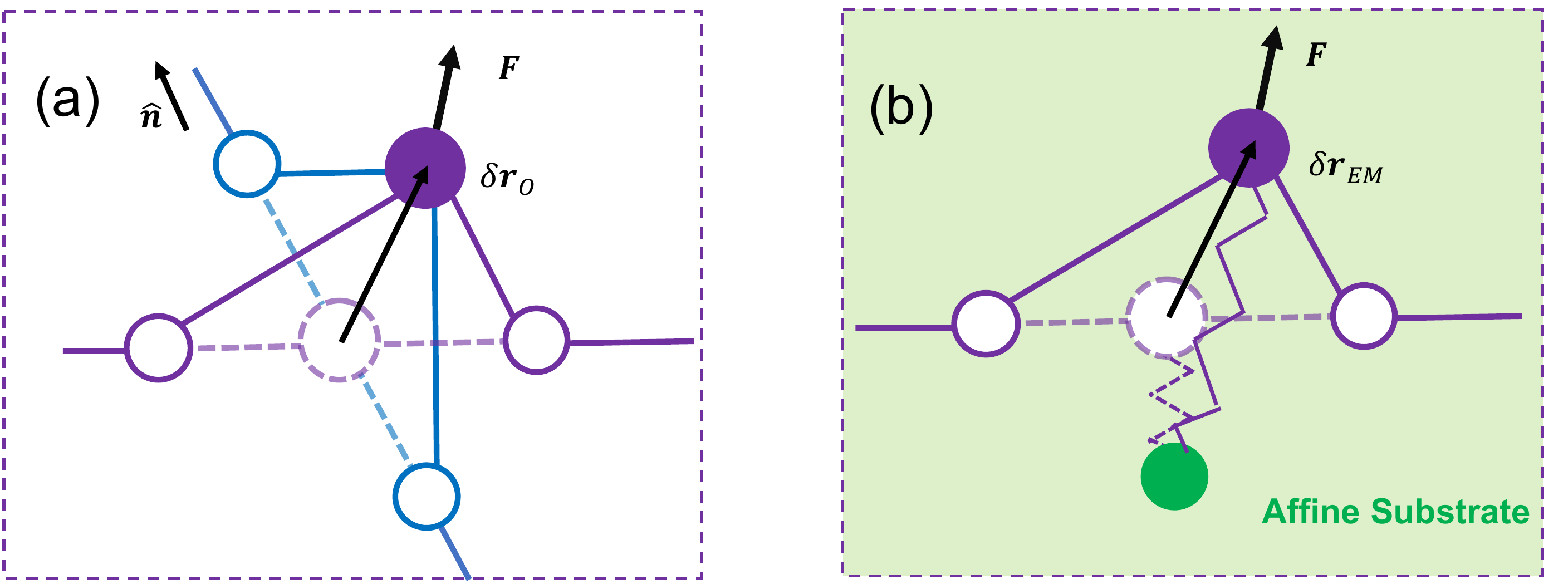}
	\caption{Sketch of the test force approach. A particular node on the purple polymer is deformed by a test force $\bm F$. The resulting displacement is $\delta \bm r_O$ in the original network (a), and $\delta \bm r_{_{\rm EM}}$ in the EMT (b). For 3D networks the adjacent nodes are assumed to be fixed, while for 2D networks the displacement of adjacent nodes need to be considered, see Fig. 4(a). 
	}
	\label{Fig.2}
\end{figure}

We start with the first requirement. Equation~(\ref{e5a}) describes the local displacement caused by the stress, which can thus be considered as a local compliance. As the stress can be decomposed to local forces on each node in the network, we exert a test force $\bm F$ on a particular node on the same polymer in both the original network and the EMT, and measure the resulting displacements, $\delta{\bm r}_O$ and $\delta{\bm r}_{_{\rm EM}}$~(see Fig.~\ref{Fig.2}). By letting $\langle \delta{\bm r}_O\rangle_{\hat{\bm n}}=\delta{\bm r}_{_{\rm EM}}$, where $\hat{\bm n}$ is the orientation of the other polymer crosslinked to the node in the original network, we obtain the values of the two spring constants, which  for 3D networks read (see Sec. I of \cite{Supp}):
 \begin{align}
K_{\perp}=K_{\parallel}=\frac{18\kappa}{\ell_c^3}.\label{e6}
 \end{align}
The equality of $K_\perp$ and $K_\parallel$ is consistent with an isotropic effective medium. Importantly, however, the node compliance is still highly anisotropic due to $H_s$. Note that in deriving Eq.~(\ref{e6}) we assumed for simplicity that all polymers are straight in the undeformed state of the original network. This assumption may not hold in real networks but is consistent with previous lattice-based simulations~\cite{stenull2011simple,Broedersz2012}. We discuss this further in Sec. IC of \cite{Supp}. 

To solve Eq.~(\ref{e5b}), one needs to find the relation between the macroscopic deformation $\uubar {\bm \Lambda}$ and the microscopic deformations ${\bm u}^{\alpha}$. This is simple in the affine limit, as noted above. For non-affine deformations the situation is more complex. To address this, instead of determining ${\bm u}^{\alpha}$ from $\uubar{\bm\Lambda}$, we do it inversely by determining $\uubar{\bm\Lambda}$  from ${\bm u}^{\alpha}$. Generally,  $\uubar{\bm\Lambda}$ is a functional of all microscopic deformations, $\uubar{\bm \Lambda}\left[{\bm u}^{1}(s),{\bm u}^{2}(s),...,{\bm u}^{N}(s)\right]$. In the small strain limit, we can always perform a linear expansion,
\begin{equation}
\begin{aligned}
\uubar{\bm \Lambda}=\sum_\alpha \int_{-L/2}^{L/2} \diff s\,{{\bm u}^\alpha(s)\cdot{\uuubar {\bm T}}^{\alpha}(s) },
\end{aligned}
\label{e7}
\end{equation} 
where ${\uuubar{\bm T}}^{\alpha}(s)$ is a third-order coefficient tensor. We find that $\uuubar{\bm T}^{\alpha}(s)$ can be uniquely determined from  three conditions: (i) the affine deformation should satisfy Eq.~(\ref{e7}), as it is a special case of the non-affine deformation; (ii) we assume the network is homogeneous on large scale, so all polymers are identical to each other except for their different orientations, leading to $\uuubar{\bm T}^{\alpha}(s)=\uuubar{\bm T}(\hat{\bm n}^{\alpha},s)$; (iii) we assume the network is isotropic~\footnote{In principle the assumptions of homogeneity and isotropy are not required for the model. Here they are adopted for simplicity. }. The full derivation of $\uuubar{\bm T}$ is detailed in Sec.~II of \cite{Supp}. A similar macroscopic-microscopic relation can be defined for the EMT as well with a coefficient tensor $\uuubar{\bm T}_{_{\rm EM}}^{\alpha}(s)$, whose value is related to $\uuubar{\bm T}^{\alpha}(s)$ via Eq.~(\ref{e5b}). For 3D networks $\uuubar{\bm T}_{_{\rm EM}}=\uuubar{\bm T}$, while for 2D networks $\uuubar{\bm T}_{_{\rm EM}}$ becomes more complicated due to the {\it floppy mode} deformation~\cite{Heussinger2006}, see discussion later. 

\begin{figure}[t]
	\centering
	\includegraphics[scale=0.4]{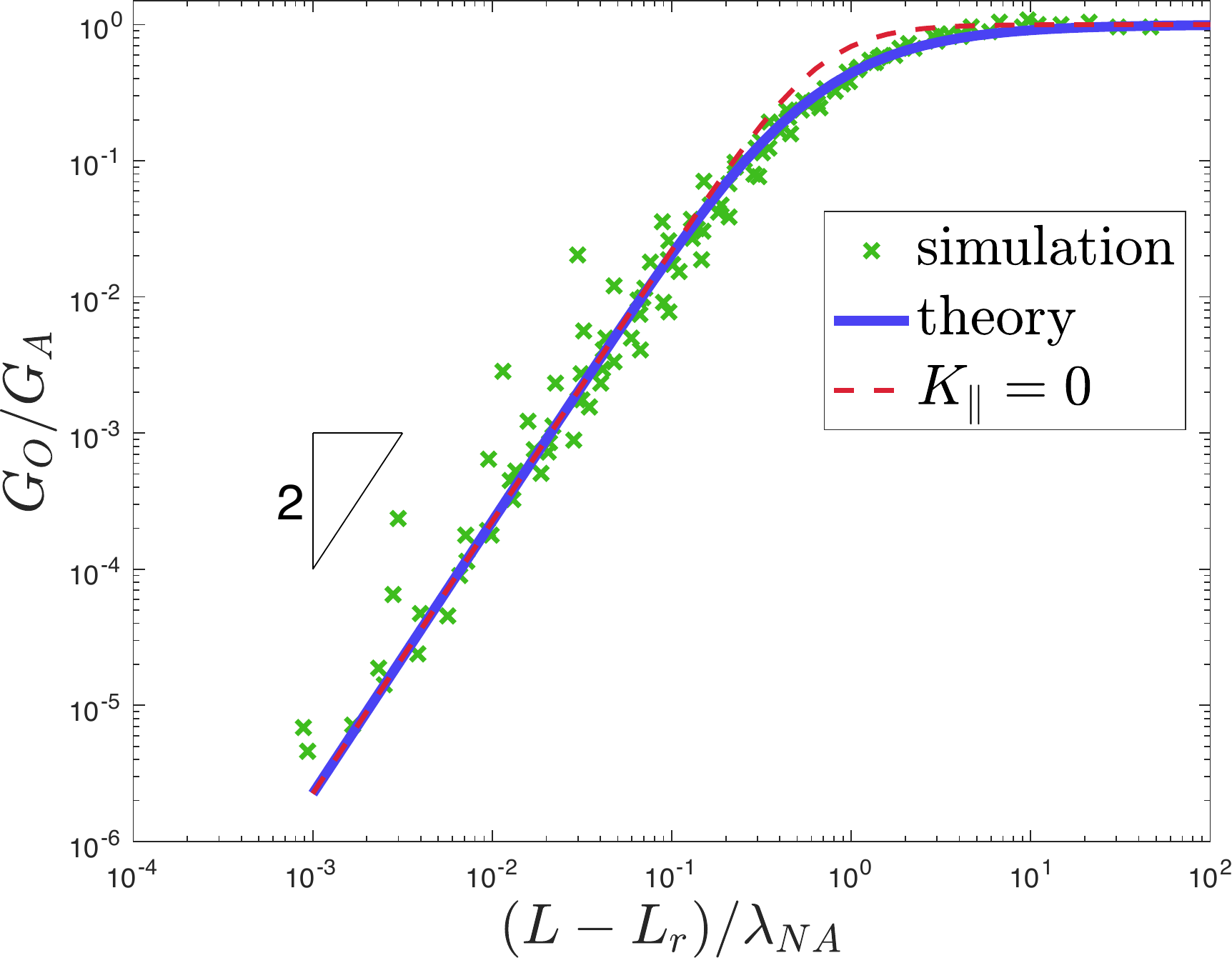}
	\caption{Shear modulus for 3D athermal networks. Simulation results of phantom-fcc-lattice network is reproduced from Ref.~\cite{Broedersz2012}, with filament length corrected by the minimum length of rigidity percolation, $L_r = 2.85\ell_c$. Theoretical prediction is plotted using Eq.~(S41) in Ref.~\cite{Supp}, which is similar to Eq.~(\ref{e8}) but calculated for networks with exponential  length distribution as in the simulation. 
	}
	\label{Fig.3}
\end{figure}

By solving Eqs.~(\ref{e5a}) and (\ref{e5b}), we have linked the EMT to the original network. The EMT elasticity $G_{_{\rm EM}}$ can be found by minimizing Eq.~(\ref{e2}) under an applied stress, which should be consistent with the elasticity of the original network $G_O$ (see Sec.~III A of Ref.~\cite{Supp} for details). For 3D athermal monodispersed networks (all polymers have the same length) we find that
\begin{equation}
\begin{aligned}
\frac{G_O}{G_A}=\left[1+\frac{4\sqrt 2 \lambda_{\rm NA}}{L}\cdot\coth\left(\frac{3 L}{\sqrt{2}\lambda_{\rm NA}}\right)\right]^{-1},
\end{aligned}
\label{e8}
\end{equation} 
where $G_A=\rho\mu/15$ is the affine linear elastic modulus, $\rho$ is the polymer length density and $\lambda_{\rm NA} = \ell_c^2/\sqrt{\kappa/\mu}$ is a characteristic non-affine lengthscale. We compare this theoretical prediction with previous simulations on lattice-based 3D networks~\cite{Broedersz2012} and find good quantitative agreement in both the scaling for small polymer length $L$ ($G_O\sim L^2$) and in the transition to an affine deformation regime for larger $L$ (Fig.~\ref{Fig.3}).  Interestingly, the non-affinity in the EMT is dominated by $K_{\perp}$. For comparison we also plot in Fig.~\ref{Fig.3} the predicted modulus with $K_{\parallel}=0$, showing a minor difference in the non-affine/affine transition region. This shows that the longitudinal deformation of the polymers is dominated by their own stretching rigidity, while $K_{\parallel}$ has a minor effect, mainly in the non-affine/affine transition where almost all longitudinal deformation is achieved from bending surrounding polymers. For simplicity we neglect $K_\parallel$ hereafter. 

Having verified our EMT using previous simulations on athermal networks, we consider thermal networks for which simulations are challenging computationally. Such challenge is due to the thermal fluctuations of the network state around its ground state, which are crucial to the elasticity of cytoskeletal networks~\cite{Gardel2004,Storm2005191}. The elasticity can be found by calculating the average strain of the Boltzmann distribution  at finite temperature $T$ (see Sec.~III B of Ref.~\cite{Supp}):
\begin{equation}
\begin{aligned}
{G_{O}}=\frac{\rho \mu_{\rm{ph}}}{15}\left(1+266.7\ell_c\ell_p/L^2\right)^{-1},
\end{aligned}
\label{e9}
\end{equation}
where $\ell_p=\kappa/(k_B T)$ is the persistence length, with $k_B$ being the Boltzmann constant. Here $\mu_{\rm ph}=100\kappa\ell_p/\ell_c^3$ is the effective stretch rigidity in the presence of thermal fluctuations. Interestingly, the limit $L\rightarrow\infty$ corresponds to a high molecular-weight analog of a phantom network, including node fluctuations. This slightly differs ($\sim 10\%$) from the limit of affinely deforming nodes with only transverse bending fluctuations \cite{MacKintosh1995,Storm2005191,Supp}. For finite $L$ we predict a strong $L$-dependence of the network elasticity that has not been identified by previous studies. Moreover, the non-affinity leads to a crucial correction to the non-linear stiffening effect~\cite{Gardel2004,Storm2005191}, as will be detailed in future work~\cite{unpublished}. 

\begin{figure}[t]
	\centering
	\includegraphics[scale=0.4]{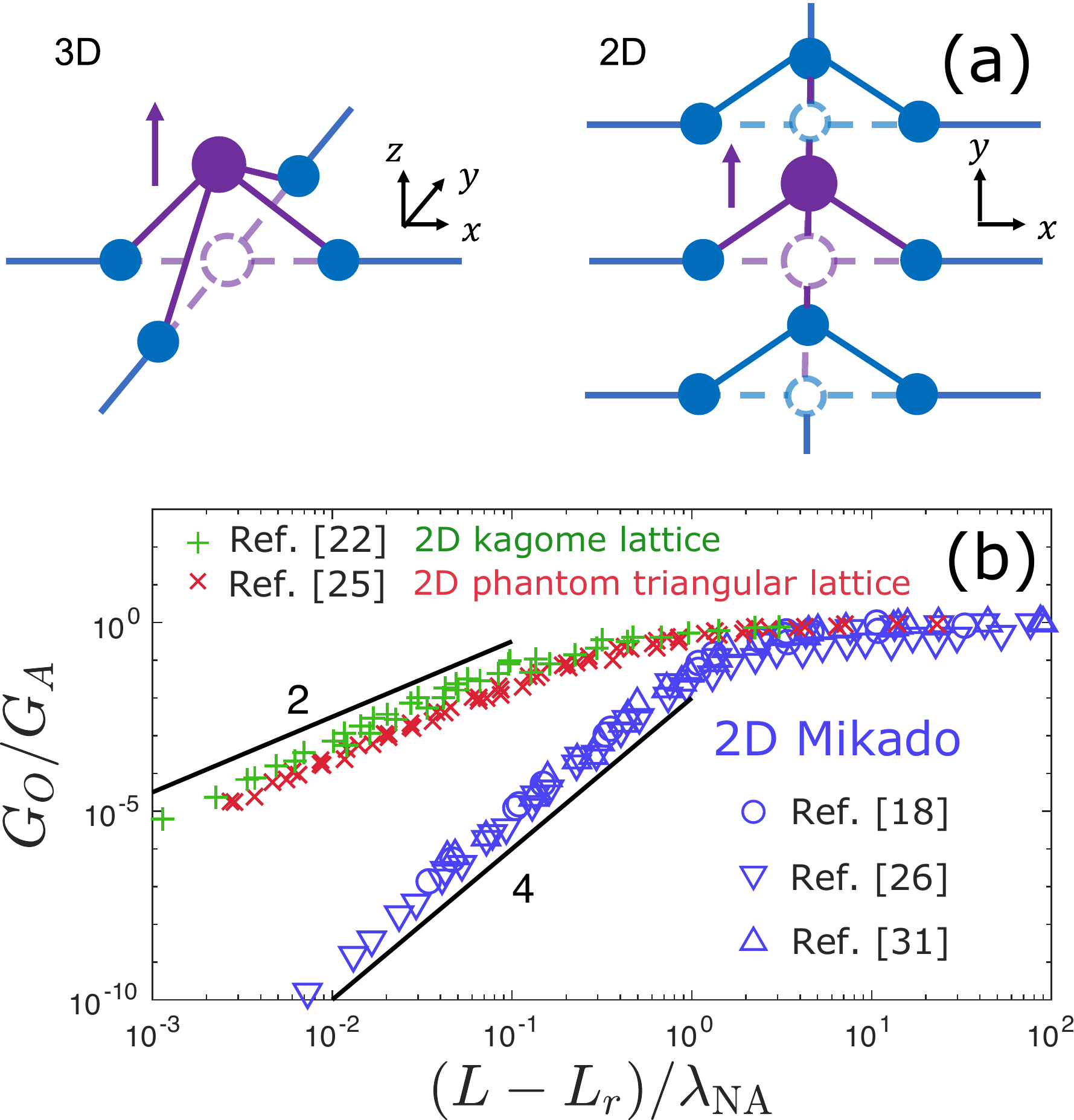}
	\caption{(a). Difference between 3D and 2D networks. In 3D networks, a crosslinker can deform in the direction perpendicular to its two connected polymers, without deforming other crosslinkers. In 2D networks, an entire polymer has to move together with the crosslinker, leading to deformation of $L/\ell_c$ crosslinkers. (b). Scaling dependence in 2D Mikado and 2D lattice-based networks. $\lambda_{\rm NA}=\kappa^{-1/4}\mu^{1/4}\ell_c^{3/2}$ for Mikado and $\lambda_{\rm NA}=\kappa^{-1/2}\mu^{1/2}\ell_c^2$ for lattice-based. $L_r=5.9\ell_c$ (Mikado), $2.94 \ell_c$ (phantom triangular) and $2.53 \ell_c$ (Kagome) are the minimum lengths for rigidity percolation. Simulation data reproduced from Ref.~\cite{Head2003PRE,Wihelm2003,Baumgarten2021} (Mikado), Ref.~\cite{Mao2013} (Kagome lattice) and Ref.~\cite{Licup2016} (phantom triangular lattice). The slight difference between Ref.~\cite{Mao2013} and Ref.~\cite{Licup2016} is due to their different lattice structures. 
	}
	\label{Fig.4}
\end{figure}

Above we have focused on 3D networks, but our theory is general to other dimensionalities. There is, however, an essential difference between 3D and 2D networks, due to the Maxwell isostatic condition for rigidity percolation for coordination number $z=Z_c =2d$ in $d$ dimensions~\cite{Maxwell1864}. For networks formed by long polymers the connectivity approaches $4$ from below. The local, near isostatic connectivity in 2D leads to long-range floppy modes~\cite{Heussinger2006,Zhou2018} that are absent in 3D, for which there is always a local floppy mode (see Fig.~\ref{Fig.4}(a)). In 2D networks, independent displacements of crosslinks are prohibited without stretching. In the limit of large $\mu$, when one crosslink in a 2D network is displaced, all other crosslinks on its connected polymer must deform in a particular way to avoid stretching deformation (see Fig.~\ref{Fig.4}(b)), leading to displacements of $L/\ell_c$ crosslinks. This {\it floppy-mode} deformation requires taking into account the coupled deformation of multiple crosslinkers when calculating both the medium rigidity (Eq.~(\ref{e6})) and the coefficient tensor (Eq.~(\ref{e7})). We find that $K_\perp\sim L$ for a 2D lattice. For Mikado networks, $K_\perp$ is further enhanced by the broad distribution of crosslink separations $\ell_c$ along the backbone~\cite{Heussinger2006}, resulting in $K_\perp\sim L^3$. As shown in Sec.~IV of~\cite{Supp}, we predict the following scaling dependences in the non-affine regime:
\begin{equation}
\begin{aligned}
G_O\sim\left\{ 
\begin{aligned}
&L^2 \qquad ({\rm 3D, any{\ }structure})\\
&L^2  \qquad(\rm 2D, lattice)\\
&L^4  \qquad(\rm 2D, Mikado)\\
\end{aligned}
\right.
\end{aligned} \, .
\label{e10}
\end{equation}
Equation~(\ref{e10}) agrees with previous numerical studies for 3D lattices~\cite{Broedersz2012}, 2D lattices~\cite{Mao2013,Licup2016}, and 2D Mikado networks~\cite{Head2003, Wihelm2003,Head2003PRE,Shahsavari2013, Baumgarten2021}, as shown in Fig.~\ref{Fig.4}(b). While various molecular weight scalings of 2D Mikado networks have been reported, the previous numerical studies are consistent with a common $(L-L_r)^4$ (see Sec.~IV of \cite{Supp}).  Interestingly, although the local network structure strongly affects the scaling dependence of 2D networks with different distributions of $\ell_c$, our model predicts a $L^2$ scaling that is robust for any structure, including potentially broad, randomly distributed $\ell_c$ in experimentally relevant 3D networks. Previous experimental studies on hydrogels and numerical studies on 3D Mikado-like networks are consistent with an $L^2$ dependence in 3D~\cite{Jaspers2014,Nguyen2016,islam2018effect}.

In conclusion, the model presented above constitutes a basis for understanding the linear elasticity of both thermal semiflexible polymer and athermal fiber networks in 2D and 3D, including non-affine effects. Such non-affine effects are known to be more important for such systems than for flexible polymer gels, although most prior work addressing non-affinity in such systems has been limited to simulation, particularly for 3D. As we have shown, the Maxwell isostatic condition results in an important difference between 2D and 3D networks, reinforcing the demand for a 3D theory. Our EMT approach predictions are in very good agreement with prior numerical simulations for athermal networks. In addition, we predict the elasticity of thermal networks and find an unexpectedly strong molecular weight dependence for which thermal simulations have been lacking. Our thermal results may aid ongoing experimental efforts to quantify non-affine effects, which have proven inconclusive to date in biopolymer networks.

An important feature of our theory is that the EMT is allowed to deform non-affinely, allowing us to capture accurately non-affine deformations of real networks. This also allows predictions of non-affine fluctuations including thermal fluctuations, in contrast to prior effective medium approaches. Our model can be extended to predict nonlinear elastic effects such as stress-stiffening~\cite{Gardel2004,Storm2005191}. This is possible even with our assumptions above of small displacements, in a way similar to prior theories of nonlinear semiflexible chain stretching~\cite{marko1995stretching,MacKintosh1995,Storm2005191}.
Our model can also be extended to address strain-controlled criticality that has previously been identified computationally~\cite{unpublished}. However, an important limitation of our approach is that it is a mean-field theory, and cannot be expected to predict anomalous critical exponents.  Moreover, with the Hamiltonian of Eq.~(\ref{e2}), the derivation of network dynamics is straightforward. Finally, our EMT approach is not limited to permanently-crosslinked networks, and can be applied also to transiently-crosslinked networks~\cite{Lieleg2008,Broedersz2010,Chen2021}. Interestingly, in Refs.~\cite{Morse2001,hinsch2007quantitative} an effective spring constant, which is conceptually similar to our effective medium rigidity, is estimated for a solution of entangled polymers. When combined with the present model, this suggests a possible model for entangled solutions. 

\begin{acknowledgements}
\emph{Acknowledgments}:
This work was supported in part by the National
Science Foundation Division of Materials Research
(Grant No. DMR-2224030) and the National Science
Foundation Center for Theoretical Biological Physics
(Grant No. PHY-2019745). The authors acknowledge fruitful discussion with T. Lubensky and M. Rubinstein. 
\end{acknowledgements}

\bibliography{citation}

\end{document}


\title{Supplementary Material}
\maketitle
\section{Derivation of the medium rigidity}
\label{sec1}
\subsection{Effective Spring Constants}
\label{sec1a}
In this section we derive the effective medium rigidity of a 3D network, {\it i.e.} the effective spring constants $K_{\parallel}$ and $K_\perp$. As discussed in the main text, we exert a test force $\bm F$ at a particular crosslinker on the same polymer in both the original network and the effective medium theory (EMT), and calculate the resulting displacements, $\delta {\bm r}_O$ and $\delta {\bm r}_{_{\rm EM}}$. By equating $\langle \delta {\bm r}_O\rangle_{\hat{\bm n}}$ and $\delta {\bm r}_{_{\rm EM}}$, where $\hat{\bm n}$ is the orientation of the other polymer crosslinked to the crosslinker in the original network, we find the values of $K_{\parallel}$ and $K_\perp$. In the calculation we assume that all polymers have straight conformations in the undeformed state of the original network, which may be a strong assumption for real 3D networks, see discussion in Sec.~\ref{sec1c}. Below we detail the calculation.  

\begin{figure}[h]
	\centering
	\includegraphics[scale=0.4]{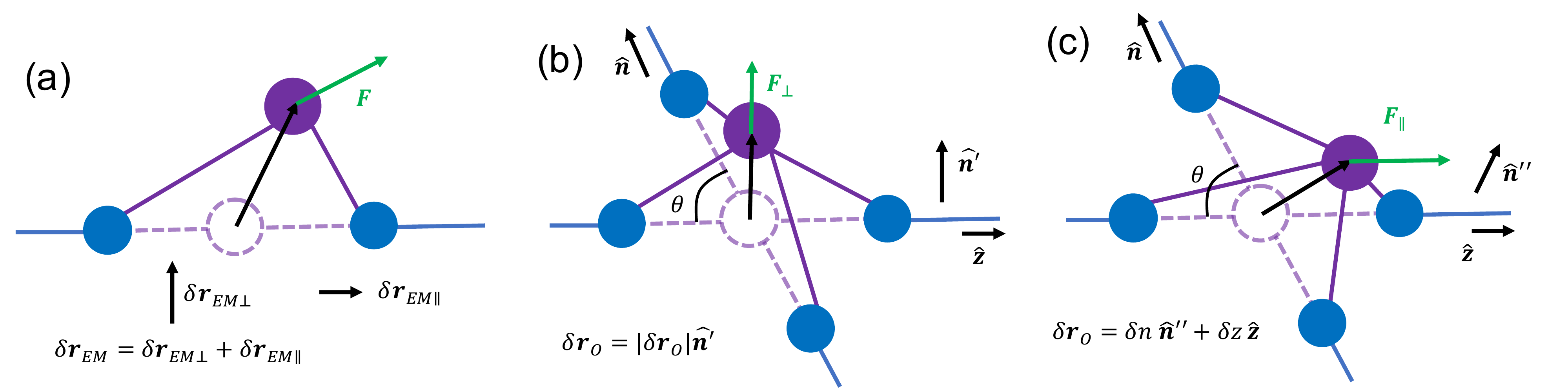}
	\caption{Illustration of the test force approach. In (a) we show the displacement of a crosslinker in the EMT due to a general test force $\bm F$. In (b) we show the displacement of a crosslinker in the original network due to a perpendicular test force $\bm F_\perp$, and in (c) we  depict the displacement of a crosslinker in the original network due to a longitudinal test force $\bm F_\parallel$. Here $\hat{\bm n}'$ is the direction that is perpendicular to both $\hat{\bm z}$ and $\hat{\bm n}$. $\hat{\bm n}'$ is a unit vector that is perpendicular to $\hat{\bm z}$ in the $\hat{\bm n}$-$\hat{\bm z}$ plane. }
	\label{Fig.S0}
\end{figure}

We start with the EMT, whose Hamiltonian is described by Eq.~(2) of the main text. Because we are mainly interested in the displacements of the crosslinkers (nodes connected to springs), we rewrite the bending and stretching energies of a particular polymer in a discrete form: 
\begin{equation}
\begin{aligned}
H_b[\bm v^\alpha(s)]&=\sum_i\frac{\kappa}{2\ell_c^3}\left|\bm v^\alpha_{\perp}(s_{i+1})-2\bm v^\alpha_{\perp}(s_{i})+\bm v^\alpha_{\perp}(s_{i-1})\right|^2\,,\\
H_s[\bm v^\alpha(s)]&=\sum_i\frac{\mu}{2\ell_c}\left|\bm v^\alpha_{\parallel}(s_{i+1})-\bm v^\alpha_{\parallel}(s_{i})\right|^2\,,
\end{aligned}
\label{S109}
\end{equation} 
where $\bm v^\alpha(s_{i})$ is the displacement of the i-th crosslinker and $\bm v^\alpha_{\parallel}(s_{i})$ and $\bm v^\alpha_{\perp}(s_{i})$ are its longitudinal and transverse components with respect to the polymer. We consider the case in which a particular crosslinker on this polymer deforms with displacement $\delta {\bm r}_{_{\rm EM}}$, while the positions of other crosslinkers are assumed to be fixed, see Fig.~\ref{Fig.S0} (a) (such assumption is only appropriate for 3D networks, not for 2D networks, see Sec.~\ref{sec4}). The resulting change in energy is calculated from Eq.~(\ref{S109}):
\begin{equation}
\begin{aligned}
\Delta H_{_{\rm EM}} = \Delta H_b(\delta {\bm r}_{\rm EM\perp}) + \Delta H_s(\delta { \bm r}_{\rm EM\parallel}) + \Delta H_K(\delta {\bm r}_{_{\rm EM}})\,,
\end{aligned}
\label{S101}
\end{equation} 
where $\Delta H_b(\delta {\bm r}_{\rm EM\perp})= (3\kappa/\ell_c^3)|\delta { \bm r}_{\rm EM\perp}|^2 $, $\Delta H_s(\delta {\bm r}_{\rm EM\parallel}) =(\mu/\ell_c)|\delta {\bm r}_{\rm EM\parallel}|^2 $ and $\Delta H_K(\delta {\bm r}_{_{\rm EM}}) = (K_\perp/2)|\delta {\bm r}_{\rm EM\perp}|^2+(K_\parallel/2)|\delta {\bm r}_{\rm EM\parallel}|^2$ are the bending, stretching and spring energy, respectively. Here $\delta {\bm r}_{\rm EM\perp}$ and $\delta {\bm r}_{\rm EM\parallel}$ are the transverse and parallel components of $\delta {\bm r}_{_{\rm EM}}$ with respect to the polymer.  Minimizing the total energy difference  $E_{_{\rm EM}}= \Delta H_{_{\rm EM}}-{\bm F}\cdot \delta {\bm r}_{_{\rm EM}}$, where ${\bm F}\cdot \delta {\bm r}_{_{\rm EM}}$ is the work done by the external force, we obtain
\begin{equation}
\begin{aligned}
\delta {\bm r}_{\rm EM\parallel} &=\frac{\bm F_{\parallel}}{2\mu/\ell_c+K_\parallel}\,,\\
\delta {\bm r}_{\rm EM\perp} &=\frac{\bm F_{\perp}}{6\kappa/\ell^3_c+K_\perp} \,,
\end{aligned}
\label{S102}
\end{equation} 
where ${\bm F}_\perp$ and ${\bm F}_\parallel$ are the transverse and parallel components of $\bm F$. 

In the original network, when a particular crosslinker on one polymer is deformed, the other polymer connected to the crosslinker is deformed at the same time. Without loss of generality, let the orientation of the first polymer be $\hat{\bm z}$, the orientation of ${\bm F_\perp}$ be $\hat{\bm x}$ and the orientation of the second polymer be $\hat{\bm n}=(\sin(\theta)\cos(\phi),\sin(\theta)\sin(\phi),\cos(\theta))$. When the crosslinker position is deformed by $\delta{\bm r}_O$, the resulting energy is the sum of bending and stretching energies of the two polymers (the bending and stretching energies are written in the discrete forms as in Eq.~(\ref{S109})):
\begin{equation}
\begin{aligned}
\Delta H_{O} &= \Delta H_b(|\delta {\bm r}_{ O}-(\delta {\bm r}_{ O}\cdot {\hat{\bm z}})\hat{\bm z}|) + \Delta H_s(\delta {\bm r}_{ O}\cdot {\hat{\bm z}})\\&+ \Delta H_b(|\delta {\bm r}_{ O}-(\delta {\bm r}_{ O}\cdot {\hat{\bm n}})\hat{\bm n}|) + \Delta H_s(\delta {\bm r}_{O}\cdot {\hat{\bm n}})\,. 
\end{aligned}
\label{S103}
\end{equation} 
The displacement is found as for the EMT by minimizing the total energy difference $E_{O}= \Delta H_{O}-{\bm F}\cdot \delta {\bm r}_{O}$ with respect to $\delta {\bm r}_{O}$.   However, an exact solution is complicated because of the coupling between the two polymers.  We consider a strong stretching limit, $\mu\gg\kappa /\ell_c^2$, which is usually valid for semiflexible biopolymers. In this case, the stretching modulus is large enough such that the polymers tend to avoid stretching deformations. Therefore, in the following calculation we only keep the leading term in $\kappa /(\mu \ell_c^2)$. To further simplify the calculation we decompose the force $\bm{F}$ into ${\bm F}_\perp$ and ${\bm F}_\parallel$ (transverse and parallel components with respect to the polymer orientation), and calculate their resulting displacements individually. Here we define the displacement caused by $\bm{F}_\perp$ as $\delta \bm{r}_{O\perp}$ and the displacement caused by $\bm{F}_\parallel$ as $\delta \bm{r}_{O\parallel}$ (note that $\delta \bm{r}_{O\perp}$ and $\delta \bm{r}_{O\parallel}$ may not be perpendicular or parallel to the polymer orientation). In the linear regime, the displacements caused by ${\bm F}_\perp$ and ${\bm F}_\parallel$ are additive, {i.e.,} $\delta {\bm r}_{O}=\delta \bm{r}_{O\perp}+\delta \bm{r}_{O\parallel}$. 

In general the displacement $\delta {\bm r}_{O\perp}$ caused by $\bm{F}_\perp$  can be aligned in any direction. However, to avoid stretching deformation, the only possible direction of $\delta {\bm r}_{O\perp}$ is the direction $\bm{\hat n}'$ that is perpendicular to both $\hat{\bm z}$ and $\hat {\bm n}$, $\hat{\bm n}'=(\sin(\phi),-\cos(\phi),0)$, see Fig.~\ref{Fig.S0} (b). Writing $\delta {\bm r}_{O\perp}=|\delta {\bm r}_{O\perp}|\hat{\bm n}'$, we have $\Delta H_O = (6\kappa/\ell_c^3)|\delta {\bm r}_{O\perp}|^2$, and
\begin{equation}
\begin{aligned}
\delta {\bm r}_{O\perp} = \frac{|\bm F_\perp|\sin(\phi)\hat{\bm n}'}{12\kappa/\ell_c^3}\,. 
\end{aligned}
\label{S104}
\end{equation} 

Because the the parallel displacement of the crosslinker must be associated with stretching deformations, the latter clearly cannot be neglected. Since $\bm F_\parallel$ is defined to be in the $\hat{\bm z}$ direction, its resulting displacement $\delta \bm{r}_{O\parallel}$ must be in  $\hat{\bm n}-\hat{\bm z}$ plane. Let $\delta \bm r_\parallel=\delta z \hat{\bm z}+\delta n \hat{\bm n}''$, where $\hat{\bm n}''=(\cos(\phi),\sin(\phi),0)$ and  $\hat{\bm z}$ form a set of orthogonal basis that spans the $\hat{\bm n}-\hat{\bm z}$ plane, see Fig.~\ref{Fig.S0} (c). Substituting $\delta \bm{r}_{O\parallel}$ in Eq.~(\ref{S103}) gives 
\begin{equation}
\begin{aligned}
&\Delta H_O=\frac{\mu}{\ell_c}\delta z^2+\frac{\mu}{\ell_c}(\delta z \cos(\theta)+\delta n\sin(\theta))^2+\frac{3\kappa}{\ell_c^3}\delta n^2+\frac{3\kappa}{\ell_c^3}(\delta n \cos(\theta)-\delta z\sin(\theta))^2\,.
\end{aligned}
\label{S105}
\end{equation}
Minimizing the total energy difference $E_O=\Delta H_O-|\bm F_\parallel|\delta z$ with respect to both $\delta z$ and $\delta n$ leads to
\begin{equation}
\begin{aligned}
\delta \bm r_{O\parallel}=\delta z \hat{\bm z} +\delta n \hat{\bm n}''=\frac{\bm F_{\parallel}}{2\mu/\ell_c+6\kappa/\ell_c^3\left[\cot(\theta)^2+1/\sin(\theta)^2\right]}+|\bm F_{\parallel}|g(\theta)\hat{\bm n}''\,, 
\end{aligned}
\label{S106}
\end{equation} 
where in the first term of Eq.~(\ref{S106}) we only keep the leading order in $\kappa/(\mu\ell_c^2)$ in the denominator. 
Here, $g(\theta)$ is some function of $\theta$, whose exact form is unimportant because the second term disappears after taking average over $\hat{\bm n}$.  

To account for the randomness in the  crosslinking angle $\hat{\bm n}$, we average  the $\delta \bm{r}_{O\parallel,\perp}$ in Eqs.~(\ref{S104},\ref{S106}) with respect to $\hat{\bm n}$, which gives (again keeping the leading order of $\kappa/(\mu\ell_c^2)$ in the denominator)
\begin{equation}
\begin{aligned}
\langle\delta \bm r_{O\parallel} \rangle_{\hat{\bm n}}&=\frac{\bm F_{\parallel}}{2\mu/\ell_c+18 \kappa/\ell_c^3}\,,\\
\langle\delta \bm r_{O\perp} \rangle_{\hat{\bm n}}&=\frac{\bm F_{\perp}}{24\kappa/\ell_c^3}\,.
\end{aligned}
\label{S107}
\end{equation} 
Here, the averages are defined by 
\begin{equation}
\begin{aligned}
\langle X \rangle_{\bm{\hat n}} = \int \diff \bm{\hat n} \cdot \bm P(\bm{\hat n})X\,,
\end{aligned}
\label{S111}
\end{equation} 
where $\bm P(\bm{\hat n})$ is the probability distribution of the orientation $\bm{\hat n}$. For the spehrical coordinate used above we have
\begin{equation}
\begin{aligned}
\langle X \rangle_{\bm{\hat n}} = \int_0^\pi \diff\theta \int_0^{2\pi} \diff \phi P(\theta)P(\phi)X\,,
\end{aligned}
\label{S110}
\end{equation} 
with $P(\theta) = (2/\pi) \sin^2(\theta)$ and $P(\phi)=1/(2\pi)$ being the distribution of angles $\theta$ and $\phi$. Note that although $P(\theta)$ shows a quadratic dependence on $\sin\theta$, it does reflect the distribution of the crosslinking angle $\theta$ in an isotropic network: one $\sin(\theta)$ factor comes from the spherical coordinate, the other originates from the fact two polymers are more likely to be crosslinked with large crosslinking angle~\cite{Heussinger2006}. For example, two parallel polymers ($\theta=0$) never crosslink, suggesting that the distribution of $\bm{\hat n}$ is not isotropic. However, this `anisotropic' distribution of $\bm{\hat n}$ is because we fix the orientation of the first polymer to be $\bm{\hat z}$, while the distribution of the orientations of all polymers is still isotropic. 

Equating Eq.~(\ref{S107}) to Eq.~(\ref{S102}) gives the two spring constants:
\begin{equation}
\begin{aligned}
K_\parallel =  \frac{18\kappa} {\ell_c^3}\,,\\
K_\perp =  \frac{18\kappa}{\ell_c^3}\,. 
\end{aligned}
\label{S108}
\end{equation} 
Equation (\ref{S108}) describes the effective medium rigidity of 3D networks with monodispersed crosslinking distance $\ell_c$. For networks with a distribution of $\ell_c$, $P(\ell_c)$, one can repeat the above calculation with an average of Eqs.~(\ref{S107}, \ref{S102}) with respect to $P(\ell_c)$, leading to
 \begin{equation}
 \begin{aligned}
 K_\parallel =  \frac{18\kappa} {\langle\ell_c^3 \rangle},\\
 K_\perp =  \frac{18\kappa}{\langle \ell_c^3\rangle}\,. 
 \end{aligned}
 \label{S112}
 \end{equation} 
 
Note that above calculation is valid for 3D networks only. In 2D networks the rigidity is different because of the 'floppy-mode' deformation, see Sec.~\ref{sec4} for details.

\subsection{Relation to Coherent Potential Approximation (CPA)}
The coherent potential approximation (CPA) was first proposed in the context of condensed matter physics~\cite{klauder1961modification} and later applied in classic effective medium theories of 2D lattice-based networks~\cite{Feng1985,Das2007,Mao20132}. The basic idea of this technique is that in order to find an effective medium for a disordered material, one needs to find the Green's functions, ${\cal G}_O$ and ${\cal G}_{_{\rm EM}}$, for the material and the effective medium, respectively.  The effective medium is then determined by letting $\langle {\cal G}_O\rangle={\cal G}_{_{\rm EM}}$, where the bracket is average over disorder realizations. This method is essentially equivalent to our derivation in Sec.~\ref{sec1a} as we show below. 

For our networks we can define the Green's functions, $\uubar{\cal G}_O(\alpha,\beta,i,j)$ and $\uubar{\cal G}_{_{\rm EM}}(\alpha,\beta,i,j)$, for the original network and the EMT, respectively. Each Green's function describes the corresponding displacement of crosslinker $i$ on polymer $\alpha$ due to a force exerted on crosslinker $j$ on polymer $\beta$. Assuming the crosslinker displacements are localized, we have 
 \begin{equation}
\begin{aligned}
\uubar{\cal G}_O(\alpha,\beta,i,j)&=\frac{\partial \delta {\bm r}_O}{\partial {\bm F}}\delta_{\alpha\beta}\delta_{ij}\,,\\
\uubar{\cal G}_{_{\rm EM}}(\alpha,\beta,i,j)&=\frac{\partial \delta {\bm r}_{_{\rm EM}}}{\partial {\bm F}}\delta_{\alpha\beta}\delta_{ij}\,. 
\end{aligned}
\label{S113}
\end{equation} 
Note that the defination of Eq.~(S113) is equivalent to $\uubar{\cal G}=(\partial^2 H/{\partial \delta {\bm r}^2})^{-1}$, which is used in some literatures~\cite{Mao20132}. Therefore, our requirement that $\langle \delta {\bm r}_O\rangle_{\hat{\bm n}}=\delta {\bm r}_{_{\rm EM}}$ naturally leads to $\langle \uubar{\cal G}_O\rangle=\uubar{\cal G}_{_{\rm EM}}$. 

\subsection{Straight-polymer assumption}
\label{sec1c}
As we have mentioned in the beginning of Sec.~\ref{sec1a}, all polymers in the original network are assumed to be straight in the undeformed state. This assumption is valid for most of the network structures studied to date in numerical simulations, including lattice-based networks in both 2D and 3D~\cite{ Broedersz2012, Mao20132, Licup2016} and off-lattice networks in 2D (Mikado). However, for real networks in 3D this assumption may not be true: whereas two nonparallel straight lines must intersect in 2D, the probability for such crossings of fibers in 3D vanishes in the limit of low volume fraction. At finite temperature, of course, such crossings will occur with finite probability and our approach should be valid at least for high enough molecular weight. In non-lattice structures, however, athermal fibers will need to bend to form a connected 3D stucture, leading to a possible prestress in the undeformed state~\cite{huisman2008monte}. While such network geometry could affect the non-affine deformation of the network, its effect can be expected to be small for weak bends required for crosslinking long fibers. Moreover, prior computational models with straight fibers quantitatively capture the mechanical properties of real networks~\cite{Sharma2016,jansen2018role}.\\
In future work, it should be possible to extend our model to the case of  non-straight fibers. Even for fibers in the original network that are non-straight in the undeformed state, we can still use straight fibers in the EMT. In this case the test force approach (CPA) for the original network should be performed with respect to a bent polymer instead of a straight polymer, and we expect different spring constants compared to Eq.~(\ref{S112}).

\section{Relation between the macroscopic and microscopic deformations }
\subsection{The original network}
In this section we derive the coefficient tensor $\uuuline {\bm T}^{\alpha}$, which relates the macroscopic ($\uubar{\bm \Lambda}_O$) and microscopic ($\bm u^\alpha$) deformations of the original network via the following equation (Eq.~(7) of the main text): 
\begin{equation}
\begin{aligned}
\uubar{\bm \Lambda}_O&=\sum_\alpha\int \diff s\,{\bm u}^{\alpha}(s)\cdot{\uuubar {\bm T}}^{\alpha}(s)\,.
\end{aligned}
\label{S201}
\end{equation} 
Here ${\uuuline {\bm T}}^{\alpha}(s)$ is a third-order coefficient tensor,  which is a function of $s$. Because the network is assumed to be homogeneous on large scale,  the polymers are equal to each other except for their individual orientations. In this case, ${\uuuline {\bm T}}^{\alpha}(s)={\uuuline {\bm T}}(\hat{\bm n}^{\alpha},s)$, where ${\hat {\bm n}}^{\alpha}$ is the orientation of the $\alpha$-th polymer. We also assume the network to be isotropic on large scale, such that it does not have any particular direction, hence, the distribution of ${\hat {\bm n}}$ is isotropic. There are only three possible third-order tensor that can be constructed from $\hat{\bm n}$: $\hat {\bm n}\uuline {\bm I}$, $\uuline{\bm I}\hat {\bm n} $ and $\hat {\bm n}\hat {\bm n}\hat {\bm n}$, and $\uuuline {\bm T}$ must be a linear combination of the three tensors:
 \begin{equation}
\begin{aligned}
\uuuline{\bm T}(\hat {\bm n},s) =  a(s)\hat {\bm n}\hat {\bm n}\hat {\bm n}+b(s) \uuline{\bm I}\hat {\bm n}+ c(s)\hat {\bm n}\uuline {\bm I}\,,
\end{aligned}
\label{S202}
\end{equation}
with $a(s)$, $b(s)$ and $c(s)$ being coefficients that depends on the position $s$. 

Our goal is to determine the values of $a$, $b$ and $c$, such that the relation between $\uubar{\bm \Lambda}_O$ and $\bm u^\alpha$ can be obtained by substituting Eq.~(\ref{S202}) into Eq.~(\ref{S201}). To do so we use the only deformation of which the macroscopic-microscopic relation is clear, the affine deformation ${\bm u}^{\alpha}(s)=s{{\uubar {\bm\Lambda}_O}}\cdot\hat{\bm n}^{\alpha} $. Because Eq.~(\ref{S201}) holds for arbitrary deformation, the affine deformation must also satisfy Eq.~(\ref{S201}), leading to
 \begin{equation}
\begin{aligned}
\uubar{\bm \Lambda}_O &=\sum_\alpha\int \diff s \left[ a(s)s\left[(\uubar{\bm \Lambda}_O\cdot\hat {\bm n}^\alpha)\cdot\hat {\bm n}^\alpha\right]\hat {\bm n}^\alpha\hat {\bm n}^\alpha+b(s)s (\uubar{\bm \Lambda}_O\cdot\hat {\bm n}^\alpha)\hat {\bm n}^\alpha+ c(s)s\left[(\uubar{\bm \Lambda}_O\cdot\hat {\bm n}^\alpha)\cdot\hat {\bm n}^\alpha\right]\uuline{\bm I} \right] \\&=  A\left\langle\left[(\uubar{\bm \Lambda}_O\cdot\hat {\bm n})\cdot\hat {\bm n}\right]\hat {\bm n}\hat {\bm n}\right\rangle_{\hat n}+B\left\langle (\uubar{\bm \Lambda}_O\cdot\hat {\bm n})\hat {\bm n}\right\rangle_{\hat n}+ C\left\langle\left[(\uubar{\bm \Lambda}_O\cdot\hat {\bm n})\cdot\hat {\bm n}\right]\uuline{\bm I}\right\rangle_{\hat n}\,,
\end{aligned}
\label{S207}
\end{equation}
where $\{A,B,C\}=N\int \diff s\{a(s),b(s),c(s)\}s$.  In the second equality we have replaced the summation with an integral, which is valid in the large $N$ limit, with $\langle X\rangle_{\hat n} $ being defined in Eq.~(\ref{S111}). This average can be found by writing $\hat{\bm n}$ in  spherical coordinate $\hat{\bm n}=(\sin(\theta)\cos(\phi),\sin(\theta)\sin(\phi),\cos(\theta))$ and using Eq.~(\ref{S110}), but with an isotropic distribution of $\bm{\hat n}$: $P(\theta)=\sin(\theta)/2$ and $P(\phi)=1/(2\pi)$ (note that unlike Sec.~\ref{sec1}, $\theta$ here is not related to a crosslinking angle). 

Because Eq.~(\ref{S207}) should be true for arbitrary deformation tensor $\uubar{\bm \Lambda}_O$, substituting two independent $\uubar{\bm \Lambda}_O$ into Eq.~(\ref{S207}) is sufficient to determine the values of $A$, $B$ and $C$.  For a bulk expansion $\uubar{\bm \Lambda}_O=\gamma \uuline {\bm I}$, the $\hat{\bm x}\hat{\bm x}$ component of Eq.~(\ref{S207}) gives $A/3 +B/3+ C = 1$. For a uniaxial deformation in $\hat{\bm x}$, $\uubar{\bm \Lambda}_O=\gamma  \hat{\bm x}\hat{\bm x}$, the $\hat{\bm x}\hat{\bm x}$ component of Eq.~(\ref{S207}) gives $A/5 + B/3 + C/3= 1$, while its $\hat{\bm y}\hat{\bm y}$ component gives $A/15 + C/3 = 0$. Solving the above three equations leads to
\begin{equation}
\begin{aligned}
A=C=0\,;\, B=3\,.
\end{aligned}
\label{S208}
\end{equation}

\begin{figure}[h]
	\centering
	\includegraphics[scale=0.4]{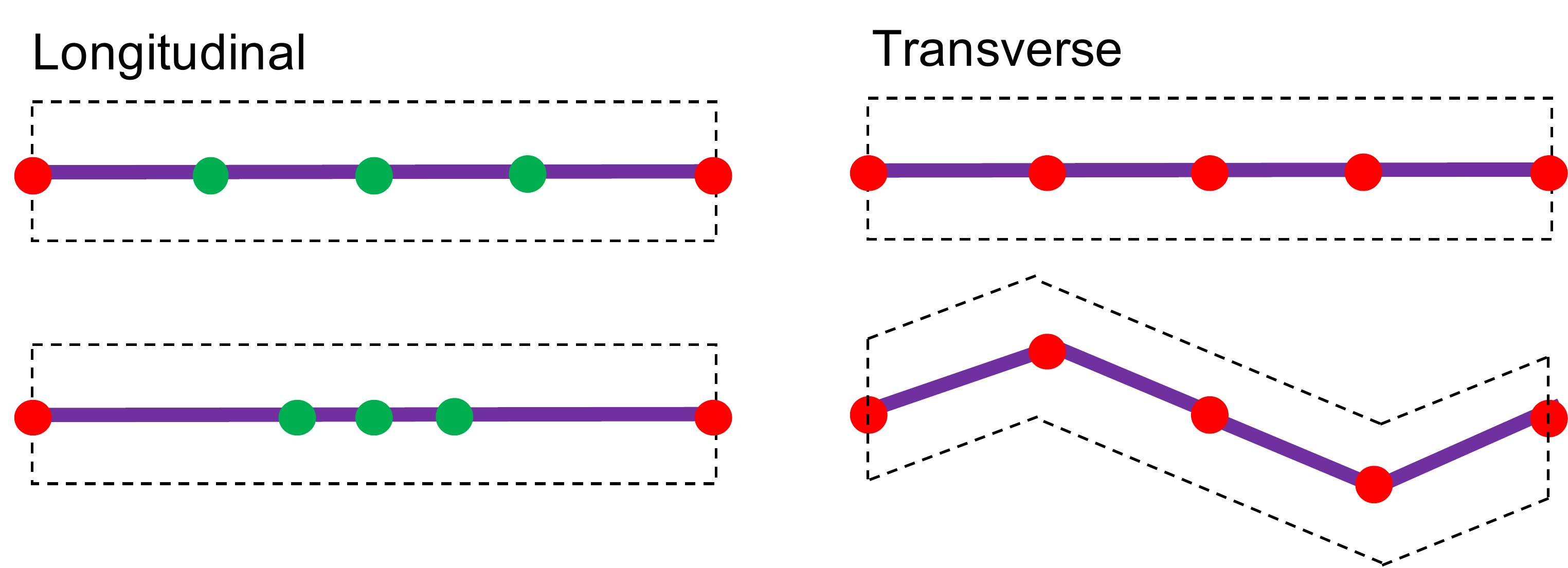}
	\caption{Illustration of the longitudinal and transverse displacements. We embed a single polymer in a box, whose shape illustrates the macroscopic deformation of the network. For longitudinal displacements (left), the macroscopic deformation is unsensative to  displacements of internal points (green dots). It is only determined by the displacements of the two ends (red dots). For transverse dispalcements (right), macroscopic deformation is affected by the displacements of all points.   }
	\label{Fig.S1}
\end{figure}

In addition to the values of $A$, $B$ and $C$, one also needs to find the dependencies of $a(s)$, $b(s)$ and $c(s)$ on $s$. Such dependencies can be found by examining the geometric properties of a single polymer. Let us use Eqs.~(\ref{S201},\ref{S202}) to calculate the deformation tensor due to a test displacement $\bm u(s)=(\bm u_{\perp}+\bm u_{\parallel})\delta(s-s_0)$ on one polymer at arbitrary position $s_0$, where $\bm u_{\perp}$ and $\bm u_{\parallel}$ are the transverse and longitudinal components of $\bm u$ with respect to $\hat{\bm n}$:
  \begin{equation}
 \begin{aligned}
 \uubar{\bm \Lambda}_O = (a(s_0)+b(s_0))(\bm u_{\parallel}\cdot{\hat {\bm n}})\hat {\bm n}\hat {\bm n}+b(s_0)\bm u_{\perp}{\hat {\bm n}}+c(s_0)(\bm u_{\parallel}\cdot{\hat {\bm n}})\uuline{\bm I}\,. 
 \end{aligned}
 \label{S203}
 \end{equation}
 The first and the last terms in Eq.~(\ref{S203}) correspond to macroscopic uniaxial and bulk expansion due to a microscopic longitudinal displacement $\bm{u}_\parallel$. As shown in Fig.~\ref{Fig.S1}, longitudinal displacements of any of the internal nodes do not affect the macroscopic deformation of the network. The only longitudinal displacement that affects the macroscopic strain is the displacement of the two polymer ends that changes its length. Writing the above statement in mathematical form, we have
\begin{equation}
\begin{aligned}
(a(s)+b(s))\sim c(s)\sim (\delta (s- L/2)-\delta(s+L/2))\,.
\end{aligned}
\label{S204}
\end{equation} 
We now switch to the second term of Eq.~(\ref{S203}), which corrsponds to a macroscopic shear deformation due to a microscopic transverse displacement $\bm{u}_\perp$. Unlike longitudinal displacements, the transverse displacements of all points of the polymer can affect the macroscopic deformation, see Fig.~\ref{Fig.S1}. We perform a trial shear deformation: opposite displacements on two points $s=s_1$ and $s=s_1+d$, $\bm u(s) = {\bm u}_\perp(\delta (s-s_1)-\delta (s-s_1-d))$. Such test displacements lead to a macroscopic deformation tensor according to Eq.~(\ref{S201}): 
\begin{equation}
\begin{aligned}
\uubar{\bm \Lambda}_O = [b(s_1)-b(s_1+d)]\bm u_{\perp}{\hat {\bm n}}\,.
\end{aligned}
\label{S205}
\end{equation} 
Because of the translational symmetry on the polymer (assuming both points are far from polymer ends), the resulting macroscopic deformation should be independent of the value of $s_1$. This suggests that $b(s)- b(s+d)=g(d)$ for arbitrary $s$ and $d$. Taking derivatives of both sides with respect to $s$, we have $b'(s)={\rm constant}$ and 
\begin{equation}
\begin{aligned}
b(s)\sim s\,.
\end{aligned}
\label{S206}
\end{equation} 
Here we have used the fact that $b(0)=0$ due to the reflectional symmetry at $s=0$ (the midpoint of the polymer). Equation~(\ref{S206}) may not be accurate when $s$ is close to the two polymer ends because the  translational symmetry no longer holds. For long polymers ($L\gg \ell_c$) such effect can be neglected. 

Combining  Eqs.~(\ref{S204}, \ref{S206}, \ref{S208}), the form of $\uuuline {\bm T}$ is finally derived :
 \begin{equation}
\begin{aligned}
\uuuline {\bm T}(\hat {\bm n},s) = [f_\parallel(s)-f_\perp(s)]\hat {\bm n}\hat {\bm n}\hat {\bm n} +f_\perp(s) \uuline {\bm I} \hat {\bm n}\,,
\end{aligned}
\label{S209}
\end{equation} 
where
 \begin{equation}
\begin{aligned}
&f_{\parallel} = a+b= \frac{3}{NL}[\delta(s-L/2)-\delta(s+L/2)]\,,\\
&f_{\perp} =b =  \frac{36}{NL^3}s\,.
\end{aligned}
\label{S210}
\end{equation} 

Above we have discussed the case in which the lengths of all polymers are the same.  If the lengths of different polymer varies, {i.e.,} the polymer length has a distribution $P(L)$, the derivation above is still valid if we replace the definitions of $\{A,B,C\}$ to 
 \begin{equation}
\begin{aligned}
\{A,B,C\} = N\left\langle\int_{-L/2}^{L/2}\diff s \,\{a(s),b(s),c(s)\}s\right\rangle_L\,,
\end{aligned}
\label{S211}
\end{equation} 
where $\langle ...\rangle_L$ is the average with respect to $P(L)$. Following the same derivation with the modified definitions of $\{A,B,C\}$, we find the modified functions $f_{\parallel}$ and $f_\perp$:

  \begin{equation}
 \begin{aligned}
 &f_{\parallel} = \frac{3}{N\langle L\rangle }[\delta(s-L/2)-\delta(s+L/2)]\,,\\
 &f_{\perp} = \frac{36}{N\langle L^3\rangle }s\,. 
 \end{aligned}
 \label{S212}
 \end{equation} 
 
Repeating the same calculation in 2D gives:

  \begin{equation}
 \begin{aligned}
 &f_{\parallel} = \frac{2}{N\langle L\rangle}[\delta(s-L/2)-\delta(s+L/2)]\,,\\
 &f_{\perp} = \frac{24}{N\langle L^3\rangle}s\,. 
 \end{aligned}
 \label{S213}
 \end{equation}

\subsection{The EMT}
Having found the macroscopic-microscopic relation for the original network, let us define a similar relation for the EMT with a coefficient tensor $\uuubar {\bm T}_{_{\rm EM}}^{\alpha}$,
\begin{equation}
\begin{aligned}
\uubar {\bm \Lambda}_{_{\rm EM}}&=\sum_\alpha\int \diff s\,{\bm v}^{\alpha}(s)\cdot\uuubar {\bm T}_{_{\rm EM}}^{\alpha}(s)\,. 
\end{aligned}
\label{S215}
\end{equation} 
The macroscopic-microscopic relations for the original network and the EMT are linked by Eq.~(5b) of the main text, {i.e.,} ${\partial \gamma_O}/{\partial \widetilde {\bm u}^{\alpha}_{i}}={\partial \gamma_{_{\rm EM}}}/{\partial \widetilde {\bm v}^{\alpha}_{i}}$, where $\widetilde{\bm u}^{\alpha}_{i}$ and $\widetilde{\bm v}^{\alpha}_{i}$ are the crosslinker displacements in the minimum energy state of the original network and the EMT, respectively. To relate Eq.~(5b) of the main text with Eqs.~(\ref{S201}, \ref{S215}), we take derivatives of the $xz$ component of Eqs.~(\ref{S201}, \ref{S215}), which gives ${\partial \gamma_O}/{\partial  {\bm u}}$ and ${\partial \gamma_{_{\rm EM}}}/{\partial {\bm v}}$. For 3D networks which are well below the isostatic point, we have ${\partial \gamma_O}/{\partial  \widetilde{\bm u}}={\partial \gamma_O}/{\partial  {\bm u}}$ and ${\partial \gamma_{_{\rm EM}}}/{\partial \widetilde{\bm v}}={\partial \gamma_{_{\rm EM}}}/{\partial {\bm v}}$. This is because in 3D networks the crosslinker displacements in the minimum energy state are localized, {i.e.,} when one crosslinker is deformed, it does not force other crosslinkers to deform together with it (see discussion in the main text). Therefore, when calculating the resulting macroscopic deformation due to a displacement of a single crosslinker (${\partial \gamma_O}/{\partial  \widetilde{\bm u}}$), that particular crosslinker would be the only one that contributes to the macroscopic deformation. Together with Eq.~(5b) of the main text this leads to $\uuubar {\bm T}_{_{\rm EM}}^{\alpha}=\uuubar {\bm T}^{\alpha}$ and 
\begin{equation}
\begin{aligned}
\uubar {\bm \Lambda}_{_{\rm EM}}&=\sum_\alpha\int \diff s\,{\bm v}^{\alpha}(s)\cdot\uuubar {\bm T}^{\alpha}(s)\,. 
\end{aligned}
\label{S214}
\end{equation} 

For 2D networks $\uuubar {\bm T}_{_{\rm EM}}$ becomes more complicated. While we still have ${\partial \gamma_{_{\rm EM}}}/{\partial \widetilde{\bm v}}={\partial \gamma_{_{\rm EM}}}/{\partial {\bm v}}$ (all polymers can deform independently in the EMT), ${\partial \gamma_O}/{\partial  \widetilde{\bm u}}\neq{\partial \gamma_O}/{\partial  {\bm u}}$ because the crosslinkers are forced to deform in {\it floppy modes}~\cite{Heussinger2006, Zhou2018}. We detail the calculation for 2D networks in Sec.~\ref{sec4}. 
\subsection{Extension for Lattice-based networks}
In this work we assume isotropic networks for simplicity, i.e., the filament orientation $\hat{\bm n}$ is isotropicly distributed. However, in numerical simulations, lattice-based network is commonly used with  discrete distribution of $\hat{\bm n}$. Our model can be extended for lattice-based networks. For this aim, when calculating the averages in Eqs.~(\ref{S107}, \ref{S207}), one should use the discrete distribution of $\hat{\bm n}$ instead of an isotropic distribution. For example, in 3D fcc lattice-based networks~\cite{Broedersz2012,Licup2016} there are 12 possible orientations (in $(\theta,\phi)$ coordinate): $(\theta_1=0,\phi_1=0)$, $(\theta_2=\pi/3,\phi_2=\pi/4)$,  $(\theta_3=\pi/3,\phi_3=3\pi/4)$, $(\theta_4=\pi/3,\phi_4=5\pi/4)$,  $(\theta_5=\pi/3,\phi_5=7\pi/4)$, $(\theta_6=\pi/2,\phi_6=0)$,  $(\theta_7=\pi/2,\phi_7=\pi)$,$(\theta_8=2\pi/3,\phi_8=\pi/4)$,  $(\theta_9=2\pi/3,\phi_9=3\pi/4)$, $(\theta_{10}=2\pi/3,\phi_{10}=5\pi/4)$,  $(\theta_{11}=2\pi/3,\phi_{11}=7\pi/4)$, $(\theta_{12}=\pi,\phi_{12}=0)$. This leads to different coefficients in Eqs.~(\ref{S108}, \ref{S210}), which further affects the elasticity of the network, compared to the isotropic prediction. For fcc-lattice simulations~\cite{Broedersz2012,Licup2016} which are compared with our analytical predictions, we find that the discrete distribution of $\hat{\bm n}$ only results in a slight difference in the elasticity ($\sim 10\%$), suggesting that the difference between fcc lattice and isotropic network is neglegible. Therefore, in the rest of the paper we assume isotropy of the network throughout. 
 
\section{Linear Elasticity of 3D Networks}
\subsection{3D Athermal Networks}
\label{sec3a}
In this section we calculate the linear elasticity of 3D athermal networks. The idea is to construct an effective medium theory, calculate its effective medium elasticity $G_{_{\rm EM}}$, and use $G_{_{\rm EM}}$ to approximate the original elasticity $G_O$. To find $G_{_{\rm EM}}$, we first need to find the minimum-energy state under an external stress $\uubar{\bm \sigma}_{_{\rm EM}}$, {i.e.,} minimizing the total energy
\begin{equation}
\begin{aligned}
E_{_{\rm EM}} = H_{_{\rm EM}}(\{\bm v^{\alpha}(s)\})- V\uubar{\bm\sigma}_{_{\rm EM}} : \uubar{\bm \Lambda}_{_{\rm EM}}\,,
\end{aligned}
\label{S301}
\end{equation} 
where $H_{_{\rm EM}}$ is defined in Eq.~(2) of the main text and $V$ is the network volume. For simplicity we set $V=1$ hereafter. 
Without loss of generality we assume the resulting deformation is a simple shear of the $xz$ plane in the $\hat{\bm x}$ direction, 
\begin{equation}
\begin{aligned}
\uubar{\bm \Lambda}_{_{\rm EM}}=
\begin{pmatrix}
0& 0& \gamma_{_{\rm EM}}\\
0 &0& 0\\
0 &0& 0
\end{pmatrix}.
\end{aligned}
\label{S327}
\end{equation} 
In the linear regime, the corresponding stress tensor is also limited to the $xz$ direction:
\begin{equation}
\begin{aligned}
\uubar{\bm \sigma}_{_{\rm EM}}=
\begin{pmatrix}
0& 0& \sigma_{_{\rm EM}}\\
0 &0& 0\\
\sigma_{_{\rm EM}} &0& 0
\end{pmatrix}.
\end{aligned}
\label{S328}
\end{equation} 
With Eqs.~(\ref{S327},\ref{S328}), we rewrite Eq.~(\ref{S301}) as
\begin{equation}
\begin{aligned}
E_{_{\rm EM}} = H_{_{\rm EM}}(\{\bm v^{\alpha}(s)\})- \sigma_{_{\rm EM}} \gamma_{_{\rm EM}}\,. 
\end{aligned}
\label{S329}
\end{equation}

 According to Eq.~(\ref{S214}), the shear strain $\gamma_{_{\rm EM}}$, which is the $xz$ component of the deformation tensor, is related to the microscopic deformations via 
  \begin{equation}
 \begin{aligned}
\gamma_{_{\rm EM}}&=\sum_\alpha\int \diff s\,{\bm v}^{\alpha}(s)\cdot\left[{\uuuline {\bm T}}^{\alpha}(s)\cddot \bm{\hat x}\bm{\hat z}\right]\,, 
 \end{aligned}
 \label{S331}
 \end{equation} 
 where $\uuuline {\bm T}$ is written in Eq.~(\ref{S209}).  To simplify the calculation, let us define  ${\bm w}^{\alpha} = {\bm v}_{\rm NA \perp}^{\alpha}+{\bm v}_{\parallel}^{\alpha}$. Substituting it in Eq.~(\ref{S331}), we have
  \begin{equation}
\begin{aligned}
\gamma_{_{\rm EM}}&=\sum_\alpha\int \diff s\,{\bm w}^{\alpha}(s)\cdot\left[{\uuuline {\bm T}}^{\alpha}(s)\cddot \bm{\hat x}\bm{\hat z}\right]+\frac{4}{5}\gamma_{_{\rm EM}}\,,
\end{aligned}
\label{S330}
\end{equation} 
where we use the isotropic distribution of the orientation $\bm{\hat n}$ (see discussion after Eq.~(\ref{S207})). 
Equation (\ref{S330}) suggests a relation between $\gamma_{_{\rm EM}}$ and $\bm{w}^\alpha$: 
\begin{equation}
\begin{aligned}
\gamma_{_{\rm EM}} &= \sum_{\alpha} \int\diff s {\bm t}^{\alpha} \cdot {\bm w}^{\alpha}\\
&= \sum_{\alpha} \int\diff s {\bm t}_\parallel^{\alpha} \cdot {\bm w}_\parallel^{\alpha}+\sum_{\alpha} \int\diff s {\bm t}_\perp^{\alpha} \cdot {\bm w}_\perp^{\alpha}\,,
\end{aligned}
\label{S302}
\end{equation} 
where ${\bm t}^{\alpha} \equiv 5{\uuuline {\bm T}}^{\alpha}\cddot (\bm{\hat x}\bm{\hat z})= {\bm t}^{\alpha}_\parallel + {\bm t}^{\alpha}_\perp$, with ${\bm t}^{\alpha}_\parallel = 5 f_\parallel n_x^{\alpha}n_z^{\alpha}\hat{\bm n}^{\alpha}$ and ${\bm t}^{\alpha}_\perp =5f_{\perp} (n_z^{\alpha}\hat{\bm x}-n_x^{\alpha}n_z^{\alpha}\hat{\bm n}^{\alpha} ) $ being the parallel and transverse components with respect to $\hat {\bm n}^{\alpha}$. We proceed by rewriting $H_{_{\rm EM}}$ of Eq.~(2) of the main text in terms of $\bm w^{\alpha} $:
\begin{equation}
\begin{aligned}
H_{_{\rm EM}}=\sum_{\alpha}\int\diff s\left[\frac{\kappa}{2}\left|\frac{\partial^2 {\bm w}^{\alpha}_\perp}{\partial s^2}\right|^2+\frac{\mu}{2}\left|\frac{\partial {\bm w}^{\alpha}_\parallel}{\partial s}\right|^2+\frac{K_\perp}{2\ell_c}\left|{ {\bm w}^{\alpha}_\perp}\right|^2+\frac{K_\parallel}{2\ell_c}\left|{ {\bm w}^{\alpha}_\parallel-s\gamma_{_{\rm EM}} n_x^{\alpha}n_z^{\alpha}\hat{\bm n}^{\alpha}}\right|^2\right]\,.
\end{aligned}
\label{S303}
\end{equation}
Note that in Eq.~(\ref{S303}) we have replaced the summation over crosslinkers with an integral. This approximation of the continuum limit is valid when $L\gg\ell_c$, which is usually true for biopolymer networks. The minimum-energy state is defined as the state in which $\delta E_{_{\rm EM}}/\delta {\bm w}^{\alpha}=0$, which is equivalent to
\begin{equation}
\begin{aligned}
\frac{\delta H_{_{\rm EM}}}{{\delta {\bm w}^\alpha_\perp}}=\sigma_{_{\rm EM}} \frac{\delta \gamma_{_{\rm EM}}}{{\delta {\bm w}^\alpha_\perp}}\,,\\
\frac{\delta H_{_{\rm EM}}}{{\delta {\bm w}^\alpha_\parallel}}=\sigma_{_{\rm EM}} \frac{\delta \gamma_{_{\rm EM}}}{{\delta {\bm w}^\alpha_\parallel}}\,. 
\end{aligned}
\label{S304}
\end{equation} 

Substituting Eqs.~(\ref{S302}, \ref{S303}) into Eq.~(\ref{S304}) leads to the following differential equations
\begin{equation}
\begin{aligned}
&\kappa\frac{\diff^4{\bm w}^{\alpha}_{\perp}}{\diff s^4}+ \frac{K_{\perp}}{\ell_c}{\bm w}^\alpha_{\perp} = \sigma_{_{\rm EM}} {\bm t}^{\alpha}_\perp\,,\\
&\mu\frac{\diff^2{\bm w}^{\alpha}_{\parallel}}{\diff s^2}+ \frac{K_{\parallel}}{\ell_c}({\bm w}^{\alpha}_{\parallel}-\gamma_{_{\rm EM}} s n_x^{\alpha}n_z^{\alpha}\hat {\bm n}^{\alpha})=\sigma_{_{\rm EM}} {\bm t}^{\alpha}_\parallel\,,
\end{aligned}
\label{S305}
\end{equation} 
together with natural boundary conditions (assuming the boundary points are free to move, {i.e.,} $\delta \bm{w}$ can take any value at the boundaries)
\begin{equation}
\begin{aligned}
&\frac{\diff^2{\bm w}^{\alpha}_{\perp}}{\diff s^2}=\frac{\diff^3{\bm w}^{\alpha}_{\perp}}{\diff s^3}=0 \qquad&(s=\pm L/2)\,,\\
&\frac{\diff{\bm w}^{\alpha}_{\parallel}}{\diff s}=\frac{15\sigma_{_{\rm EM}}n_x^{\alpha}n_z^{\alpha}\hat{\bm n}^{\alpha}}{N\langle L \rangle\mu } \qquad&(s=\pm L/2)\,.  
\end{aligned}
\label{S332}
\end{equation} 
In the second line of Eq.~(\ref{S332}) we have integrated the second line of Eq.~(\ref{S305}) around the boundaries. The solution of Eqs.~(\ref{S305},~\ref{S332}) is
\begin{equation}
\begin{aligned}
{\bm w}^{\alpha}_{\perp} &= \frac{10 \ell^4_c\sigma_{_{\rm EM}} s}{N\langle L^3\rangle \kappa} \,(n_z^{\alpha}\hat{\bm x}-n_x^{\alpha}n_z^{\alpha}\hat{\bm n}^{\alpha})\,,\\
{\bm w}^{\alpha}_\parallel&=\left[\gamma_{_{\rm EM}} s + \left(\frac{15\sigma_{_{\rm EM}}}{N\langle L \rangle} -\mu\gamma_{_{\rm EM}}\right)\cdot \frac {\sinh(\epsilon s)} {\epsilon \mu \cosh(\epsilon L/2)}\right]n_x^{\alpha}n_z^{\alpha}\hat {\bm n}^{\alpha}\,. 
\end{aligned}
\label{S306}
\end{equation} 
Here $\epsilon =3\sqrt 2 /\lambda_{\rm NA}$ with $\lambda_{\rm NA}=\ell_c^2/\sqrt{\kappa/\mu} $ being a characteristic length of non-affine deformation. In Eq.~(\ref{S306}) we have used Eq.~(\ref{S212}) for the values of $f_\parallel$ and $f_\perp$ and Eq.~(\ref{S108}). Substituting the solution of Eq.~(\ref{S306}) back into Eq.~(\ref{S302}) gives:
 \begin{equation}
 \begin{aligned}
\gamma_{_{\rm EM}} &= \sigma_{_{\rm EM}}\left(\frac{15}{N\mu\langle L \rangle}+\frac{20\ell_c^4\epsilon\langle L \rangle }{N\langle L^3\rangle\kappa\langle\tanh(\epsilon L/2)\rangle}\right)\,. 
 \end{aligned}
 \label{S308}
 \end{equation}
 The shear elasticity, defined as $G_O= G_{_{\rm EM}}=\sigma_{_{\rm EM}}/\gamma_{_{\rm EM}}$, is
  \begin{equation}
\begin{aligned}
\frac {G_{O}}{G_A} &= \left[1+\frac{4\sqrt 2 \lambda_{\rm NA}\langle L \rangle^2 }{\langle L^3\rangle\langle\tanh(3\sqrt 2 L/2\lambda_{\rm NA})\rangle}\right]^{-1}\,,
\end{aligned}
\label{S309}
\end{equation}
where $G_A=\rho\mu/15$ is the affine modulus. Here $\rho=N\langle L\rangle$ is the polymer length density ($V=1$). Equation~(\ref{S309}) is a general result for any polymer length distribution. For monodispersed networks, $P(L) = \delta (L-\langle L\rangle)$, we have
  \begin{equation}
\begin{aligned}
\frac {G_{O}}{G_A} &= \left[1+\frac{4\sqrt 2  }{(L/\lambda_{\rm NA})\tanh(3\sqrt 2 L/2\lambda_{\rm NA})}\right]^{-1}\,.
\end{aligned}
\label{S310}
\end{equation}
To compare with the simulation results in Ref.~\cite{Broedersz2012}, we also consider
networks with  exponentially distributed polymer length, $P(L)=\exp(-L/\langle L\rangle)/\langle L\rangle$. In this case, we have $\langle L^3\rangle=6\langle L \rangle^3$, and
  \begin{equation}
\begin{aligned}
\frac {G_{O}}{G_A} &= \left[1+\frac{2\sqrt 2 \lambda_{\rm NA} }{3\langle L\rangle/\lambda_{\rm NA}\langle\tanh(3\sqrt 2 L/2\lambda_{\rm NA})\rangle}\right]^{-1}\,.
\end{aligned}
\label{S311}
\end{equation}

To examine the importance of $K_{\parallel}$ for non-affine deformations, we set the value of $K_\parallel$ to zero when solving Eq.~(\ref{S305}), leading to 
\begin{equation}
\begin{aligned}
{\bm w}^{\alpha}_\parallel&=\frac{15\sigma_{_{\rm EM}}}{N\langle L\rangle\mu}sn_x^{\alpha}n_z^{\alpha}\hat {\bm n}^{\alpha}\,,
\end{aligned}
\label{S325}
\end{equation} 
where the solution of ${\bm w}_\perp$ is as in Eq.~(\ref{S306}). The corresponding $G_O$ for $K_\parallel=0$ is
  \begin{equation}
\begin{aligned}
\frac {G_{O}}{G_A} &= \left[1+\frac{8 \lambda_{\rm NA}^2\langle L \rangle }{3\langle L^3\rangle}\right]^{-1}\,.
\end{aligned}
\label{S326}
\end{equation}
We find that removing $K_\parallel$ almost does not change the shear modulus, see Fig.~3 of the main text. The reason for this is that the longitudinal displacement is always restricted by $\mu$, even in the absence of  $K_\parallel$. This is different from $K_\perp$, which when setting to zero results in diverging ${\bm w}_\perp$ and therefore vanishing $G$ (see Eq.~(\ref{S306})). Therefore, we conclude that $K_\perp$ is crucial for calculating the non-affine deformation while $K_\parallel$ is not. 

\subsection{3D Thermal Networks}
\label{sec3b}
So far we have found non-affine deformation for athermal netowrks ($T=0$), 
in this section we consider thermal networks ($T>0$).  The non-zero temperature generates thermal fluctuations of the polymer displacements ${\bm w}^{\alpha}$. Therefore, the network shear deformation $\gamma_{_{\rm EM}}$ is fluctuating and the shear modulus is now defined as $G_{_{\rm EM}}=\sigma_{_{\rm EM}}/\langle\gamma_{_{\rm EM}} \rangle$, where the bracket stands for average over noise realizations. $\gamma_{_{\rm EM}}$ is related to ${\bm w}^{\alpha}$ via Eq.~(\ref{S302}):
\begin{equation}
\begin{aligned}
\gamma_{_{\rm EM}} 
&=\frac{15}{N\langle L\rangle}\sum_{\alpha} n_x^{\alpha}n_z^{\alpha} \hat{\bm n}^\alpha\cdot\left[\bm w_\parallel^{\alpha}(s=L/2)-\bm w_\parallel^{\alpha}(s=-L/2)\right]+\sum_{\alpha} \int\diff s\, {\bm t_\perp}^{\alpha} \cdot {\bm w}_\perp^{\alpha}\,,
\end{aligned}
\label{S320}
\end{equation} 
where we have used values of $f_\parallel$ in Eq.~(\ref{S212}). 
 For simplicity hereafter we consider polymers in the  inextensible limit ($\mu\to\infty$), which is appropriate for most biopolymers. In this case the network deformation can be described by only the transverse displacements $\bm w^{\alpha}_\perp$, while the longitudinal end-to-end distance is related to the transverse displacements via~\cite{Broedersz2014}
  \begin{equation}
\begin{aligned}
\hat{\bm n}^\alpha\cdot\left[\bm w_\parallel^{\alpha}(s=L/2)-\bm w_\parallel^{\alpha}(s=-L/2)\right]=L_0-\frac 1 2\int \diff s \left|\frac{\partial \bm w^{\alpha}_\perp}{\partial s}\right|^2\,,
\end{aligned}
\label{S312}
\end{equation}
with $\ell_p=\kappa/k_B T$ being the persistence length. Here $L_0$ is the contraction of the end-to-end distance without external stress, which will be defined below in Eq.~(\ref{S317}) . 

For inextensible polymers with $K_\parallel =0$ (here $K_\parallel$ is set to zero for simplicity, since it only slightly affects the non-affinity of athermal networks, see discussion after Eq.~(\ref{S326})), the network Hamiltonian is (see Eq.~(\ref{S303}))
\begin{equation}
\begin{aligned}
H_{_{\rm EM}}=\sum_{\alpha}\int\diff s\left[\frac{\kappa}{2}\left|\frac{\partial^4 {\bm w}^{\alpha}_\perp}{\partial s^4}\right|^2+\frac{K_\perp}{2\ell_c}\left|{ {\bm w}^{\alpha}_\perp}\right|^2\right]\,. 
\end{aligned}
\label{S321}
\end{equation}
Because the Hamiltonian is quadratic, the thermal average $\langle\bm w^{\alpha}_\perp\rangle$ is same as the minimum-energy solution of Eq.~(\ref{S306}). The fluctuations around the minimum-energy state, $\bm w^{\alpha}_\perp-\langle\bm w^{\alpha}_\perp\rangle$, is expanded into Fourier series:
  \begin{equation}
\begin{aligned}
\bm w^{\alpha}_\perp-\langle\bm w^{\alpha}_\perp\rangle = \sum_q \Bigg[w^{\alpha}_{1q}\sin\Big(q(s+L/2)\Big) \hat{\bm n}_1^{\alpha}+w^{\alpha}_{2q}\sin\Big(q(s+L/2)\Big) \hat{\bm n}_2^{\alpha}\Bigg]\,,
\end{aligned}
\label{S313}
\end{equation}
where $q=m\pi/L$ ($m=1,2,3...$) is the wave number, $\hat{\bm n}_1^{\alpha}$ and $\hat{\bm n}_2^{\alpha}$ are two unit vectors perpendicular to both $\hat{\bm n}^{\alpha}$ and each other, such that $\hat{\bm n}^{\alpha}$, $\hat{\bm n}_1^{\alpha}$ and $\hat{\bm n}_2^{\alpha}$ form a basis of the 3-dimensional space. Substituting Eq.~(\ref{S313}) into Eq.~(\ref{S312}) leads to
  \begin{equation}
\begin{aligned}
\left\langle\hat{\bm n}^\alpha\cdot\left[\bm w_\parallel^{\alpha}(s=L/2)-\bm w_\parallel^{\alpha}(s=-L/2)\right]\right\rangle=L_0-\frac{L}{4}\sum_qq^2\Bigg[\left\langle(w_{1q}^{\alpha})^2\right\rangle+\left\langle(w_{2q}^{\alpha})^2\right\rangle\Bigg]\,,
\end{aligned}
\label{S314}
\end{equation}
where we have neglected the contribution from $\langle\bm w^{\alpha}_\perp\rangle$, because $\langle\bm w^{\alpha}_\perp\rangle\sim\sigma_{_{\rm EM}}$, which adds a term $\sim \sigma_{_{\rm EM}}^2$ in Eq.~(\ref{S314}), hence, it does not contribute to linear elasticity. 
The thermal fluctuations also perturb the system total energy $E_{_{\rm EM}}$ from its minimum $E_{\rm min}$. Substituting Eqs.~(\ref{S320}, \ref{S314}) into Eq.~(\ref{S329}) gives
  \begin{equation}
\begin{aligned}
E_{_{\rm EM}}-E_{\rm min} = \frac{L}{4}\sum_{mq}(\kappa q^4 +\tau^{\alpha}_\parallel q^2+ \frac{K_\perp}{\ell_c})\left[(w_{1q}^{\alpha})^2+(w_{2q}^{\alpha})^2\right]\,,
\end{aligned}
\label{S315}
\end{equation}
where $\tau^{\alpha}_\parallel = 15\sigma_{_{\rm EM}} n_x^{\alpha}n_z^{\alpha}/(N\langle L\rangle)$ is the effective longitudinal tension on the $\alpha$-th polymer. The equilibrium amplitudes of the Fourier modes then satisfy the equipartition theorem, 
  \begin{equation}
\begin{aligned}
\left\langle\left(w_{1q}^{\alpha}\right)^2\right\rangle=\left\langle\left(w_{2q}^{\alpha}\right)^2\right\rangle=\frac{2k_B T}{L(\kappa q^4+ \tau^{\alpha}_\parallel q^2 +K_\perp/\ell_c)}\,.
\end{aligned}
\label{S316}
\end{equation}
Setting $\tau^\alpha_\parallel=0$ in Eq.~(\ref{S316}) gives the zero-stress amplitudes. Because $\langle \bm w\rangle$ must be zero in the absence of stress, substituting the zero-stress amplitudes into Eq.~(\ref{S314}) gives
  \begin{equation}
\begin{aligned}
L_0 = \sum_q\frac{k_B Tq^2}{\kappa q^4 +K_\perp/\ell_c}\,.
\end{aligned}
\label{S317}
\end{equation}
Using Eqs.~(\ref{S316},\ref{S317}) and Eq.~(\ref{S314}), we obtain the average longitudinal displacement
  \begin{equation}
\begin{aligned}
\left\langle\hat{\bm n}^\alpha\cdot\left[\bm w_\parallel^{\alpha}(s=L/2)-\bm w_\parallel^{\alpha}(s=-L/2)\right]\right\rangle&=\sum_q\left[\frac{k_B Tq^2}{\kappa q^4 +K_\perp/\ell_c}-\frac{k_B Tq^2}{\kappa q^4 +\tau_\parallel^{\alpha}q^2+K_\perp/\ell_c}\right]\\&=\sum_q\frac{k_B T\tau_\parallel^{\alpha}q^4}{\left(\kappa q^4 +K_\perp/\ell_c\right)^2}\,. 
\end{aligned}
\label{S318}
\end{equation}
In the second equation we have used the fact that $\sigma_{_{\rm EM}}$ is small, thus $\tau^\alpha_{\parallel}$ is also small and we expand it to linear order in $\sigma_{_{\rm EM}}$. Because $K_\perp/\ell_c\sim\kappa/\ell_c^4 \gg \kappa/L^4$, we can replace the summation in Eq.~(\ref{S318}) with an integral, leading to
  \begin{equation}
\begin{aligned}
\left\langle\hat{\bm n}^\alpha\cdot\left[\bm w_\parallel^{\alpha}(s=L/2)-\bm w_\parallel^{\alpha}(s=-L/2)\right]\right\rangle&\simeq 0.01\frac{L\ell_c^3k_B T}{\kappa^2}\tau_\parallel^{\alpha}\equiv\frac{L\tau_\parallel^{\alpha}}{\mu_{\rm ph}}\,,
\end{aligned}
\label{S319}
\end{equation}
where $\mu_{\rm ph}\simeq100\kappa \ell_p/{\ell_c^3}$ is the effective stretch rigidity of a phantom chain, which is  discussed below in Sec.~\ref{sec3c}. We then calculate the network shear deformation by substituting Eq.~(\ref{S319}) and the solution of $\langle \bm w_\perp^{\alpha}\rangle$ (Eq.~(\ref{S306})) into Eq.~(\ref{S320}):
\begin{equation}
\begin{aligned}
\langle\gamma_{_{\rm EM}}\rangle =  \sigma_{_{\rm EM}}\left(\frac{40\ell_c^4 }{N\langle L^3\rangle\kappa}+\frac{15}{N\langle L\rangle\mu_{\rm ph}}\right)\,. 
\end{aligned}
\label{S322}
\end{equation} 
Eq.~(\ref{S322})  gives the shear modulus, 
\begin{equation}
\begin{aligned}
{G_O}=  \frac{\rho\mu_{\rm ph}}{15}(1+\lambda_{\rm NA}^2\langle L\rangle/\langle L^3\rangle)^{-1}\,, 
\end{aligned}
\label{S324}
\end{equation} 
where $\lambda_{\rm NA}=16.3\sqrt{\ell_c\ell_p}$ is the length scale governing the non-affine to affine transition. 
Equation (\ref{S324}) depends on the polymer length distributions. For monodispersed networks we have
\begin{equation}
\begin{aligned}
{G_O}=  \frac{\rho\mu_{\rm ph}}{15}(1+\lambda_{\rm NA}^2/L^2)^{-1}\,,
\end{aligned}
\label{S333}
\end{equation} 
which is Eq.~(9) of the main text.

\subsection{High-Molecular-Weight Limit}
\label{sec3c}
Above  we have calculated the linear elasticity of athermal networks (Sec.~\ref{sec3a}) and thermal networks (Sec.~\ref{sec3b}). For both networks we find that the ratio between filament length $L$  and the non-affine length scale $\lambda_{\rm NA}$ governs non-affinity (Eqs.~(\ref{S309}, \ref{S324})). Here $\lambda_{\rm NA}=\ell_c^2/\sqrt{\kappa/\mu}$ for athermal networks and $\lambda_{\rm NA}\simeq16.3\sqrt{\ell_c\ell_p}$ for thermal networks. While for athermal networks $G_O$ reduces to  the affine modulus $G_A$ when $L\gg\lambda_{\rm NA} $, for thermal networks the affine modulus is not recovered for large $L$. In fact, taking $L\gg\lambda_{\rm NA}$ in Eq.~(\ref{S324}) we have ${G_O}= G_{\rm ph}$, where $G_{\rm ph}=\rho\mu_{\rm ph}/15$ is a phantom-network-like limit of our model. Here, $\mu_{\rm ph}\simeq100\kappa\ell_p/\ell_c^3$ is the effective stretch rigidity of a phantom chain, which is different by a numerical factor from the affine stretch modulus $\mu_{A}=90\kappa\ell_p/\ell_c^3$~\cite{Gittes1998, Morse19987030}. There are two reasons for this difference. The first reason is that in our model the crosslinkers of each polymer are allowed to deform in a non-affine manner, similar to phantom networks of flexible polymers~\cite{Flory1985}. The second reason is that our model takes into account the bending interactions between adjacent segments, which are not taken into account in the traditional affine model~\cite{Gittes1998, Morse19987030}.  Interestingly, these two factors changes the effective stretch rigidity in opposite directions: the non-affine deformation of a phantom chain lowers the stretch rigidity, while the inter-segment bending interactions strengthen it. As a result of the combining effect of the above two factors,  $\mu_{\rm ph}$ is $10\%$ larger than $\mu_A$. 

Let us also clarity the range of validity of our theory which depends on two ratios, $\ell_c/L$ and $\lambda_{\rm NA}/L$. Our predicted non-affinity is only significant when $L<\lambda_{\rm NA}$, and can thus be interpreted as a {\it finite-length correction} for large $\lambda_{\rm NA}/L$ values. On the other hand, the assumption $L\gg\ell_c$ is used throught, implying that an additional  {\it finite-length correction} should be applied for large $\ell_c/L$.  Our theory is valid in the regime $\ell_c \ll \lambda_{\rm NA}$ where one can neglect the latter finite length correction. For athermal networks this regime is equivalent to $\sqrt{\kappa/\mu}\ll \ell_c$, which is clearly satisfied considering the fact that $\sqrt{\kappa/\mu}$ is of the molecular scale~\cite{Broedersz2014}. For thermal networks our theory is valid for $\ell_c \ll\ell_p$, which is exactly the definition of semiflexible polymers. To conclude, for both athermal and thermal networks $\ell_c \ll \lambda_{\rm NA}$ is naturally satisfied.

 \section{2D Networks}
 \label{sec4}
 In the previous section we have  calculated the non-affine deformation of 3D networks. In this section we apply the theory on 2D networks and demonstrate the difference between 2D and 3D networks. 
 
\begin{figure}[t]
 	\centering
 	\includegraphics[scale=0.5]{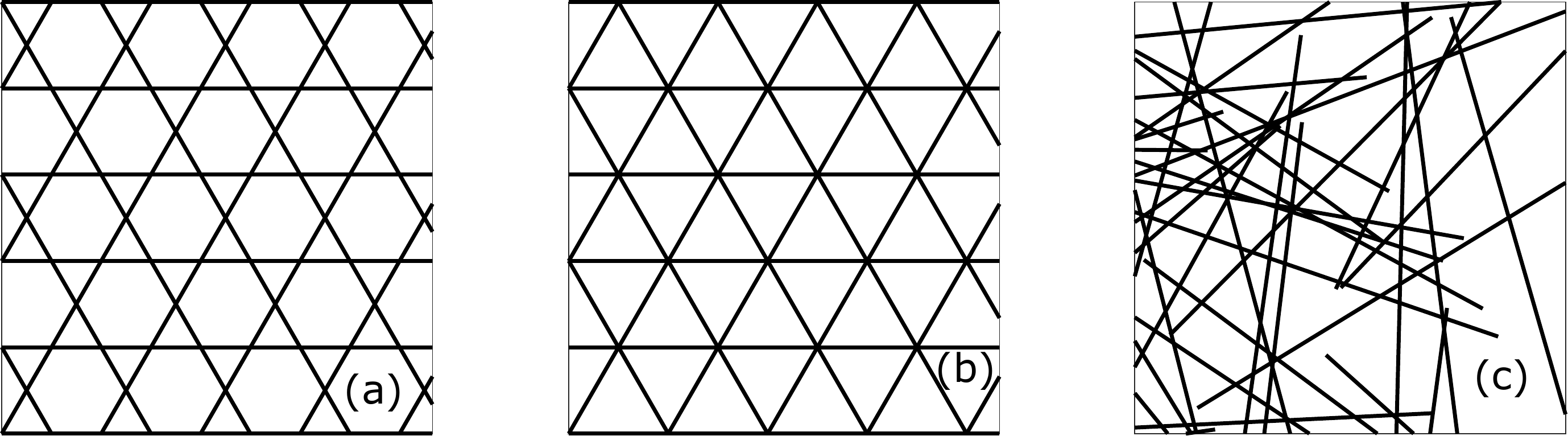}
 	\caption{Illustration of different 2D network structures. (a) Undiluted kagome lattice. Each line is an infinitely long polymer and each intersection of two polymers is a crosslink. (b) Undiluted phantom triangular lattice. While three polymers intersect at the same node, only two of them are randomly chosen to form a crosslink, and the third one just passes through the node. The two lattices are then diluted, i.e., bonds between adjacent crosslinks are randomly removed, to achieve the desired average polymer length. (c) Mikado network. Polymers with the same length are randomly distributed in a 2D plane with random orientations. Each intersection of two polymers is a crosslilnk. One can show that in Mikado networks the crosslinking length follows an exponential distribution~\cite{Heussinger2006}. } 
 	\label{Fig.S2}
 \end{figure}

 For 3D athermal networks we have shown that its linear elasticity is governed by a non-affine length $\lambda_{\rm NA}=\ell_c^2/\sqrt{\kappa/\mu}$, where in the bend-dominated regime $L\ll\lambda_{\rm NA}$ we find that $G_O\sim L^2$. Similar non-affine length, together with the scaling dependence in the bend-dominated regime, has also been observed previously in numerical simulations of 2D networks~\cite{Head2003, Wihelm2003,Head2003PRE,Mao2013,Licup2016}, although the value of the non-affine length and the scaling exponent may vary with network structure. For any network structure $G_O\sim L^\xi$ in the bend-dominated regime, with $\lambda_{\rm NA}=\ell_c^{1+2/\xi}(\kappa/\mu)^{-1/\xi}$~\cite{Broedersz2014}.  Numerical studies have revealed the scaling exponents $\xi$ in various network structures, which are $\xi=2$ in 2D lattice-based networks~\cite{Mao2013,Licup2016} and $\xi=4$ or $5$ in 2D Mikado networks~\cite{Head2003PRE,Wihelm2003} (see Fig.~\ref{Fig.S2} for details of the network structures). Although several analytical models have been proposed for some of the  above network structures, a unified model which explains how the exponent $\xi$ changes with network structure is lacking. Below we provide an analytical understanding of 2D networks using our EMT approach. 
 
 Before we detail the derivation, let us point out an important difference between 2D and 3D networks, which is the Maxwell connectivity for rigidity percolation. Maxwell found that a critical connectivity (isostatic point) for spring networks is $Z=2d$, where $d$ is the dimension~\cite{Maxwell1864}. This suggests that $Z=4$ in 2D and $Z=6$ in 3D. If the connectivity exceeds this critical connectivity, a spring network without bending rigidity would be rigid, while subisostatic networks are floppy without bending rigidity.  In crosslinked networks, a crosslinker that connects two polymers usually provides a local connectivity of $4$ ($3$ if the crosslinker is near one of the polymer ends). Therefore, 2D networks are very close to their isostatic point, while 3D networks are far from it. As we discuss in the main text, the result is that in 3D networks a crosslinker can have a localized displacement, while in 2D networks the displacement must be non-localized. In fact, in 2D networks the total number of independent deformation modes equals the number of polymers $N$~\cite{Zhou2018}. Each of these deformation modes, the so-called `floppy modes', corresponds to the deformation of an entire polymer~\cite{Heussinger2006}. Such non-local displacements in 2D networks will affect both the spring constant $K_\perp$ and the coefficient tensor $\uuubar T_{_{\rm EM}}$ in the EMT, as we discuss in detail below. 
 
 \subsection{Effective Medium Rigidity in 2D EMT} 
 Because we are interested in the scaling exponent $\xi$ in the bend-dominated regime, for simplicity we assume $\mu\to \infty$ heretofore. In this case the longitudinal displacements ${\bm u}_\parallel$ can be neglected, such that the only spring constant that affects the network deformation is $K_\perp$. 
 
 We first consider a lattice-based network, {i.e.,} with a constant crosslinking length $\ell_c$. The Hamiltonians of the original network and the EMT are
  \begin{equation}
  \begin{aligned}
H_O &= \sum_{\alpha}\int\diff s\,\frac{\kappa}{2}\left|\frac{\partial^2 {\bm u}^{\alpha}_\perp}{\partial s^2}\right|^2\,,\\H_{_{\rm EM}}&=\sum_{\alpha}\int\diff s\,\left[\frac{\kappa}{2}\left|\frac{\partial^2{\bm v}^{\alpha}_\perp}{\partial s^2}\right|^2+\frac{K_\perp}{2\ell_c}\left|{ {\bm v}^{\alpha}_\perp}\right|^2\right]\,.
  \end{aligned}
  \label{S401}
  \end{equation} 
We follow the same steps as in Sec.~\ref{sec1}, {i.e.,} calculating the resulting displacement $\delta \bm r$ of a test force $\bm F$ in both networks. Similar to what we found in Sec.~\ref{sec1}, in the original network the displacement $\delta \bm r_O$ of one crosslinker leads to a bending energy $\Delta H_b\sim (\kappa/\ell_c^3)|\delta \bm r_O|^2$. Following the introductory paragraph and Ref.~\cite{Heussinger2006}, in 2D networks when one crosslinker is deformed, it must deform an entire polymer together with it, leading to deformation of $L/\ell_c$ crosslinkers in total. Therefore, we have $\Delta H_O\sim (L/\ell_c)(\kappa/\ell_c^3)|\delta \bm r_O|^2$, and 
  \begin{equation}
\begin{aligned}
\delta \bm r_{O\perp}\sim\frac{\bm F_\perp}{(L/\ell_c)(\kappa/\ell_c^3)}\,.
\end{aligned}
\label{S402}
\end{equation} 
Unlike in the original network, the crosslinker displacements in the EMT are localized. Therefore, $\delta \bm r_{\rm EM\perp}$ still follows Eq.~(\ref{S102}). Equating Eq.~(\ref{S102}) with Eq.~(\ref{S402}) gives
   \begin{equation}
 \begin{aligned}
K_\perp\sim\frac{L}{\ell_c}\frac{\kappa}{\ell_c^3}\,. 
 \end{aligned}
 \label{S403}
 \end{equation} 

In Mikado networks, the crosslinking distance $\ell_c$ is not a constant, but follows an exponential distribution, $P(\ell_c)=\exp(-\ell_c/\langle\ell_c\rangle)/\langle \ell_c \rangle$. To account for this distribution of $\ell_c$, one should modify the bending energy of a single crosslinker to $\Delta H_b\sim \langle\kappa/\ell_c^3\rangle|\delta \bm r_O|^2$. Together with the number of crosslinkers on each polymer $L/\langle \ell_c \rangle$, Eq.~(\ref{S403}) is modified as follows:
    \begin{equation}
 \begin{aligned}
 K_\perp\sim\frac{L}{\langle\ell_c\rangle}\left\langle\frac{\kappa}{\ell_c^3}\right\rangle\,,
 \end{aligned}
 \label{S404}
 \end{equation} 
 where the averages are with respect to $P(\ell_c)$. Notably, the average $\langle \kappa/\ell_c^3\rangle$ diverges because $\ell_c$ can be arbitrarily small. To resolve this divergence, we adopt the minimum-cutoff method proposed in Ref.~\cite{Heussinger2006}: If the distance between two crosslinkers is smaller than a cutoff distance $\ell_{\rm min}$, we assume their bending interaction vanishes. This is because when two crosslinkers are too close to each other, the energy to cause relative displacement between them is extremely large, such that it is energetically favorable to move the entire connected polymer instead of bending the two crosslinkers. The cutoff length is thus defined as 
     \begin{equation}
 \begin{aligned}
\frac{\kappa}{\ell_{\rm min}^3}=\frac{L}{\langle\ell_c\rangle}\left\langle\frac{\kappa}{\ell_c^3}\right\rangle_{\ell_c>\ell_{\rm min}}\,, 
 \end{aligned}
 \label{S405}
 \end{equation} 
 where the left hand side is the effective rigidity to introduce bending between two crosslinkers with distance $\ell_{\rm min}$, and the right hand side is the effective rigidity to deform an entire polymer. For $\ell_{\rm min}\ll \langle \ell_c \rangle$, the average of $\kappa/\ell_c^3$ is dominated by the contribution from $\ell_c\sim \ell_{\rm min}$, in which case it can be approximated by $\kappa/(\ell_{\rm min}^2\langle \ell_c\rangle)$. With this approximation, Eq.~(\ref{S405}) gives $\ell_{\rm min}\sim \langle \ell_c\rangle^2/L$, and 
     \begin{equation}
 \begin{aligned}
K_\perp\sim \frac{\kappa}{\langle\ell_c\rangle^6}L^3\,.
 \end{aligned}
 \label{S406}
 \end{equation} 

To conclude, we found that in 2D lattice-based networks $K_\perp\sim L^2$ and in 2D Mikado networks $K_\perp\sim L^3$ as we state in the main text. 

\subsection{Relate Macroscopic and Microscopic Deformations in 2D EMT}
Having found the effective medium rigidity, let us continue by deriving the coefficient tensor which relates the macroscopic and microscopic deformations in 2D EMT.  Due to the non-local displacements in 2D networks, the coefficient tensor $\uuubar {\bm T}_{_{\rm EM}}$ of the 2D EMT differs from that of the original network. Folllowing Eq.~(\ref{S215}), in 2D EMT the deformation tensor $\uubar {\bm \Lambda}_{_{\rm EM}}$ is related to the microscopic displacements $\bm v^{\alpha}(s)$ via
 \begin{equation}
\begin{aligned}
\uubar {\bm \Lambda}_{_{\rm EM}}= \sum_{\alpha}\int \diff s\,{\bm v}_\perp^{\alpha}(s)\cdot{\uuubar {\bm T}^{\alpha}_{_{\rm EM}}}(s)\,.
\end{aligned}
\label{S407}
\end{equation} 
Note that we have only included the transverse displacements, since we assume here that $\mu\to \infty$ which prohibits any longitudinal displacements. Because in 2D there is only one transverse direction, we have ${\bm v}_\perp^{\alpha}(s)=v^\alpha(s)\hat{\bm n}_\perp^\alpha$ with $\hat{\bm n}_\perp^\alpha$ being the unit vector of the transverse direction of the $\alpha$-th polymer. Taking the $xy$ component (assuming that is the shear direction) of Eq.~(\ref{S407}) and writing it in a discrete form we have
 \begin{equation}
\begin{aligned}
\gamma_{_{\rm EM}} = \sum_{\alpha i}r^\alpha_{i}v^\alpha_{i}\,,
\end{aligned}
\label{S410}
\end{equation} 
where $v^\alpha_{i} = v^{\alpha}(s_i)$ is the displacement of each crosslinker with $s_i$ being the position of the $i$-th crosslinker, and $r^\alpha_i\equiv\ell_c\hat{\bm n}_\perp^\alpha\cdot{\uuubar {\bm T}^{\alpha}_{_{\rm EM}}}(s_i):\hat{\bm x}\hat{\bm y}$. 

For the original network, we have (see Eq.~(\ref{S201}))
 \begin{equation}
\begin{aligned}
\uubar {\bm \Lambda}_{O} &= \sum_{\alpha}\int \diff s\,{\bm u}_\perp^{\alpha}(s)\cdot{\uuuline {\bm T}}^{\alpha}(s)=\sum_{\alpha}\int \diff s f_\perp(s){\bm u}_\perp^{\alpha}(s)\hat{\bm n}^{\alpha}\,,
\end{aligned}
\label{S408}
\end{equation} 
where we have used Eq.~(\ref{S209}). We also rewrite the $xy$ component of Eq.~(\ref{S408}) in the discrete version
 \begin{equation}
\begin{aligned}
\gamma_O = \sum_{\alpha i}t^\alpha_{i}u^\alpha_{i}\,,
\end{aligned}
\label{S409}
\end{equation} 
where $u^\alpha_{i} = {\bm u}_\perp^{\alpha}(s_i)\cdot\hat{\bm n}_\perp^\alpha$ and 
 \begin{equation}
\begin{aligned}
t_{i}^\alpha=\ell_c f_\perp(s_i)\hat{\bm n}_\perp^{\alpha}\hat{\bm n}^{\alpha}:\hat{\bm x}\hat{\bm y}\,. 
\end{aligned}
\label{S421}
\end{equation} 

We are interested in finding an EMT that satisfies $\partial \gamma_O/\partial \tilde { u}^\alpha_{i}=\partial \gamma_{_{\rm EM}}/\partial \tilde {v}^\alpha_{i}$, where $\tilde { u}^\alpha_{i}$ and $\tilde { v}^\alpha_{i}$ are crosslinker displacements in the minimum-energy state (see Eq.~(5b) of the main text). Because in the minimum-energy state other $M=L/\ell_c-1$ crosslinkers must deform together with the crosslinker `$i$', we can write
 \begin{equation}
\begin{aligned}
\frac{\partial \gamma_O}{\partial \tilde u^\alpha_{i}}= t^\alpha_{i} + \sum_{m=1}^{M}c(\alpha, \alpha_m,i,i_m)t^{\alpha_m}_{i_m}\,,
\end{aligned}
\label{S411}
\end{equation} 
where the summation represents the contribution from other crosslinkers in the floppy mode deformation with the $m$-the crosslinker of the floppy mode having crosslinker number $i_m$ on the $\alpha_m$-th polymer. $c(\alpha, \alpha_m,i,i_m)=\partial\tilde u_{i_m}^{\alpha_m}/\partial\tilde u_{i}^{\alpha}$ is a dimensionless coefficient which is determined by the crosslinking angle. 
In the EMT, on the other hand, the crosslinker displacements are uncorrelated as in 3D, such that
 \begin{equation}
\begin{aligned}
\frac{\partial \gamma_{_{\rm EM}}}{\partial \tilde v^{\alpha}_{i}}= r^{\alpha}_{i}\,,
\end{aligned}
\label{S412}
\end{equation} 
Comparing Eq.~(\ref{S412}) and Eq.~(\ref{S411}) leads to 
 \begin{equation}
\begin{aligned}
r^{\alpha}_{i}=t^\alpha_{i} +\sum_{m=1}^{M}c(\alpha, \alpha_m,i,i_m)t^{\alpha_m}_{i_m}\,.
\end{aligned}
\label{S413}
\end{equation} 
To write it in a more general way, 
 \begin{equation}
\begin{aligned}
r^{\alpha}_{i}\simeq\sum_{\beta, j}c(\alpha,\beta,i,j)t^\beta_j\,, 
\end{aligned}
\label{S420}
\end{equation} 
where $c(\alpha,\beta,i,j)\neq0$ if $u^\alpha_i$ and $u^\beta_j$ are in the same floppy mode, and $c(\alpha,\beta,i,j)=0$ in other cases. Here we have neglected the first term in Eq.~(\ref{S413}) as the summation term dominates when $M\gg 1$. 

\subsection{Linear Elasticity in 2D networks}
Having obtained both the effective medium rigidity and the coefficient tensor for 2D EMT, we now calculate the linear elasticity of the networks. Because in 2D networks $K_\perp$ shows linear (lattice) or cubic (Mikado) dependence on $L$, in the $L\gg\ell_c $ limit we have $K_\perp\gg\kappa /\ell_c^3$, such that the spring energy dominates $H_{_{\rm EM}}$ and  the bending term in $H_{_{\rm EM}}$ can be neglected (see Eq.~(\ref{S401})). We then write $H_{_{\rm EM}}$ in the discrete form as
 \begin{equation}
\begin{aligned}
H_{_{\rm EM}} \simeq \sum^\alpha_{i}\frac{K_\perp}{2}(v^\alpha_{i})^2\,. 
\end{aligned}
\label{S414}
\end{equation} 
Minimizing the total energy $E_{_{\rm EM}}=H_{_{\rm EM}}-\gamma_{_{\rm EM}}\sigma_{_{\rm EM}}$ leads to 
 \begin{equation}
\begin{aligned}
v^\alpha_{i} = \sigma_{_{\rm EM}}r^\alpha_{i}/K_\perp \,,
\end{aligned}
\label{S415}
\end{equation} 
and substituting Eq.~(\ref{S415}) back into Eq.~(\ref{S410}) yields
 \begin{equation}
\begin{aligned}
\gamma_{_{\rm EM}}&= \frac{\sigma_{_{\rm EM}}}{K_\perp}\sum_{\alpha i }(r^\alpha_{i})^2 \\&=\frac{\sigma_{_{\rm EM}}}{K_\perp}\sum_{\beta \beta' j' j' }\sum_{\alpha \alpha' i i' } c(\alpha, \beta,i,j)c(\alpha', \beta',i',j')t^{\beta}_{j}t^{\beta'}_{j'} \,,
\end{aligned}
\label{S416}
\end{equation} 
where we have used Eq.~(\ref{S420}). Because the crosslinking angles are randomly distributed, the coefficients $c(\alpha, \beta,i,j)$ can be regarded as a random variable with zero average and $\langle c(\alpha, \beta,i,j)c(\alpha', \beta',i',j') \rangle=\langle c^2(\alpha, \beta,i,j)\rangle\delta_{\alpha \alpha'}\delta_{\beta \beta'}\delta_{i i'} \delta_{j j'}$. Here $\langle c^2(\alpha, \beta,i,j)\rangle=\bar c^2$ if $u^\alpha_i$ and $u^\beta_j$ are in the same floppy mode, and zero in other cases. With that we have
 \begin{equation}
\begin{aligned}
\sum_{\alpha \alpha' i i' } c(\alpha, \beta,i,j)c(\alpha', \beta',i',j') \simeq M \bar c^2\delta_{\beta \beta'}\delta_{j j'}\,, 
\end{aligned}
\label{S419}
\end{equation}
where the number $M$ emerges because there are $M$ crosslinkers in the same floppy mode with crosslinker $(\beta,j)$. 
From Eqs.~(\ref{S416},\ref{S419}) we conclude that
 \begin{equation}
\begin{aligned}
\gamma_{_{\rm EM}}\sim \frac{\sigma_{_{\rm EM}}}{K_\perp}\frac{L}{\ell_c}\sum_{\beta j}(t^\beta_{j})^2\,. 
\end{aligned}
\label{S417}
\end{equation} 
Finally we utilize Eqs.~(\ref{S213},\ref{S403}, \ref{S406}, \ref{S421}) and Eq.~(\ref{S417}) to arrive at the non-affine deformation
\begin{equation}
\begin{aligned}
\gamma_{_{\rm EM}}\sim\left\{ 
\begin{aligned}
&\sigma_{_{\rm EM}} \rho^{-1}\kappa^{-1}\ell_c^4L^{-2}  \qquad(\rm 2D, lattice-based)\\
&\sigma_{_{\rm EM}} \rho^{-1}\kappa^{-1}\langle\ell_c\rangle^6L^{-4}  \qquad(\rm 2D, Mikado)\\
\end{aligned}
\right.
\end{aligned} \,,
\label{S418}
\end{equation}
and the corresponding linear elasticity
\begin{equation}
\begin{aligned}
G_O\approx G_{_{\rm EM}}\sim\left\{ 
\begin{aligned}
&L^2  \qquad(\rm 2D, lattice-based)\\
&L^4  \qquad(\rm 2D, Mikado)\\
\end{aligned}
\right.
\end{aligned} \,.
\label{e10}
\end{equation}

The $L^2$ dependence of 2D lattice-based networks agrees with previous numerical simulations on both random-diluted triangular networks~\cite{Licup2016} and kagome lattice~\cite{Mao2013}. For Mikado networks previous numerical simulations have arrived at different conclusions: While Refs.~\cite{Wihelm2003,Heussinger2006} found an exponent close to $4$ which agrees with our prediction, Ref.~\cite{Head2003PRE} and recent studies on higher molecular weights~\cite{Shahsavari2013, Baumgarten2021} claim that the exponent is $5$. We believe that an important detail in the data processing may be the cause of this disagreement. In Ref.~\cite{Wihelm2003} the filament length $L$ (which is equivalent to the polymer length density in their original paper) is modified to $L-L_r$, where $L_r=5.9\ell_c$ is the rigidity percolation length. This modification tends to correct finite-length effects of the simulation. However, in Refs~\cite{Head2003PRE,Shahsavari2013, Baumgarten2021} such correction is not used. To test whether this correction is the origin of the different observed exponents, we replot the data obtained from Ref.~\cite{Wihelm2003,Shahsavari2013, Baumgarten2021} in Fig.~4 of the main text. We find that all data collapse on a single curve, which shows $G\sim L^4$ in the bend-dominated regime. This suggests that the exponent $5$ is an artifact the finite polymer length, and our theoretical prediction of $G\sim L^4$ agrees with multiple numerical simulations.

\bibliography{citation}